\documentclass[fleqn,usenatbib]{mnras}

\usepackage{newtxtext,newtxmath}

\usepackage{booktabs}  

\usepackage[T1]{fontenc}
\usepackage{subfigure}
\usepackage{graphicx}
\usepackage{tikz}
\usepackage{float}

\usetikzlibrary{arrows.meta, angles, quotes, decorations.markings}
\usetikzlibrary{decorations.pathreplacing}

\DeclareRobustCommand{\VAN}[3]{#2}
\let\VANthebibliography\thebibliography
\def\thebibliography{\DeclareRobustCommand{\VAN}[3]{##3}\VANthebibliography}

\usepackage{subfigure} 

\title[Three-point alignments in FLAMINGO]{Three-point intrinsic alignments of galaxies and haloes in the FLAMINGO simulations}

\author[Vedder et al.]{
Casper Vedder,$^{1}$\thanks{E-mail: vedder@strw.leidenuniv.nl}
Thomas Bakx,$^{2}$
Nora Elisa Chisari,$^{1, 2}$
Henk Hoekstra,$^{1}$
Matthieu Schaller$^{1,3}$
\\
$^{1}$Leiden Observatory, Leiden University, PO Box 9513, NL-2300 RA Leiden, The Netherlands\\
$^{2}$Institute for Theoretical Physics, Utrecht University,
Princetonplein 5, 3584 CC, Utrecht, The Netherlands \\
$^3$Lorentz Institute for Theoretical Physics, Leiden University, PO Box 9506, NL-2300 RA Leiden, the Netherlands}

\date{Accepted XXX. Received YYY; in original form ZZZ}

\pubyear{\the\year{}}

\begin{document}
\label{firstpage}
\pagerange{\pageref{firstpage}--\pageref{lastpage}}
\maketitle


\begin{abstract}
Third-order statistics provide information beyond two-point measures, but extracting this information requires accurate and consistent modelling. We measure and detect the three-point correlation function and third-order aperture mass statistics of intrinsic alignments (IA) for galaxies and for haloes with $M_{\rm halo} > 10^{13}\,{\rm M}_\odot$ in the $(2.8\,\mathrm{Gpc})^3$ simulation volume of the \textsc{FLAMINGO} hydrodynamical simulation suite. We model the third-order aperture mass statistics and show that on large scales both the galaxy and halo samples are well described by the tree-level effective field theory (EFT) of IA across the three dark matter density-shape combinations and a wide range of triangle configurations, with the alignment amplitude consistent with that inferred from two-point statistics. We compare the full EFT to several other models: a version neglecting the velocity-shear term, the non-linear alignment model, and a reduced EFT assuming co-evolution relations that follow from the assumption that alignment is linear in Lagrangian space. The first two models yield biased constraints on the alignment amplitude, but the reduced EFT performs remarkably well, achieving a low reduced chi-squared and minimal bias. We examine the redshift and mass dependence of the higher-order bias parameters, finding that the linear Lagrangian bias assumption is approximately satisfied across the explored halo mass and redshift ranges for both galaxies and haloes. These co-evolution relations can be valuable for photometric shear surveys, where limited constraining power on IA parameters favours models with fewer free parameters.
\end{abstract}

\begin{keywords}
cosmology:theory -- large-scale structure of the Universe -- galaxies:haloes
\end{keywords}


\section{Introduction}
Next-generation photometric galaxy surveys, such as those conducted by \textit{Euclid} \citep{Mellier2024}, \textit{the Nancy Grace Roman Space Telescope} \citep{spergel2015}, and the Rubin Observatory \citep{ivezic2019}, will provide high-precision shape measurements for an unprecedented number of galaxies, while stage IV spectroscopic surveys such as those conducted by \textit{Euclid} \citep{euclidnisp2025}, DESI \citep{desi2024} and \textit{SPHEREx} \citep{spherex} will deliver accurate redshifts for a vast number of galaxies. These surveys contain tremendous amounts of cosmological information, offering new opportunities for probing the large-scale structure of the Universe with more precision and at smaller scales. A well-known example of this is weak gravitational lensing: the paths of photons are perturbed by matter on their way to Earth. This results in coherent distortions in the observed shapes of distant galaxies, leading to measurable correlations that trace the underlying matter distribution (see e.g. \citet{mandelbaum2018, kilbinger2015} for reviews). This can then be used to constrain our models of the Universe. However, to extract all information from these measurements, our theoretical models must be of a similar accuracy as the precision of the measurements. 

A key challenge in interpreting weak-lensing measurements is disentangling the portion of the observed shape correlations that is caused by gravitational lensing from that arising from intrinsic galaxy alignments (IA). These intrinsic correlations, generated by the gravitational interaction between galaxies and the surrounding large-scale structure (see e.g. \citet{troxel2015,joachimi2016, chisari2025} for reviews), can significantly bias cosmological inferences if not accurately modelled \citep{krause2016}.
At the same time, IA represents a valuable cosmological and astrophysical probe in its own right. Because intrinsic alignments trace the large-scale structure in a distinct way, they encode information about the early Universe \citep{schmidt2015, chisari2016, kurita2023}, potential parity-violating physics \citep{philcox2024}, and late-time structure growth \citep{okumura2023, okumura2024}, among other phenomena \citep{orlando2020, harvey2021, vandompseler2023, saga2024}. Moreover, IA measurements can shed light on galaxy formation and evolution processes \citep{soussana2020, Bhowmick2020, bate2020, vanheukelum2025}.

Studies of galaxy-shape correlations have traditionally focused on second-order correlations between the density and shape fields, such as position–shape and shape–shape correlations. For Gaussian fields, two-point statistics capture all the available information. However, due to non-linear gravitational collapse, the matter and shape fields develop non-Gaussian features, causing information to leak into higher-order statistics. Several statistics have been proposed to probe this non-Gaussian information (see, e.g. \citet{howls2023} for an overview). These statistics hold the promise of not only containing additional information \citep{kilbinger2005, gatti2020, Burger2024}, but also showing a different dependence on the complex and non-linear astrophysics of galaxy formation \citep{Semboloni2013, Foreman2020, Broxterman2025, zhou2025, marinichenko2025}, potentially offering a way to better inform our models and extend the analyses to smaller scales. 

The three-point correlation function (3PCF), along with its Fourier-space counterpart, the bispectrum, is the lowest-order non-Gaussian statistic and is expected to contain substantial information on quasi–non-linear scales. An additional practical advantage of the 3PCF is that it can be measured directly from a shape catalogue without requiring intermediate processing steps. For several cosmological probes, it has been demonstrated that a joint analysis of the power spectrum and bispectrum yields significantly tighter parameter constraints. This improvement arises because the two statistics exhibit different parameter dependencies, which help break degeneracies while providing complementary information \citep[see, e.g.,][]{amico2024,Burger2024,des3pt2025,verdiani2025}. However, incorporating third-order statistics into a two-point analysis is non-trivial, as it requires a consistent theoretical framework in which the second- and third-order models are compatible. This requirement poses a significant theoretical challenge. 

Compared to higher-order formalisms in weak lensing and galaxy clustering, higher-order intrinsic alignments remain relatively under-explored. Early work on the modelling and mitigation of intrinsic alignments in third-order weak lensing statistics is reviewed by \citet{troxel2015}. Initial estimates of the impact of three-point intrinsic alignments on lensing analyses were presented by \citet{Semboloni2008}, followed by more detailed theoretical modelling in subsequent studies, including \citet{krause2011} and \citet{troxel2012}. Recent work has measured the position-position-shape aperture mass signal in the LOWZ sample of the Sloan Digital Sky Survey (SDSS) Baryon Oscillation Spectroscopic Survey (BOSS) \citep{linke2024}, reporting that for this sample, a relatively simple model suffices. Furthermore, the bispectrum of IA has been measured for haloes in the IllustrisTNG simulations \citep{pyne2022}. However, the latter is restricted by the relatively small box (300 Mpc) of the IllustrisTNG simulations. From a theoretical standpoint, \citet{schmitz2018} and \citet{vlah2020} derived the IA bispectrum at tree level in perturbation theory. \citet{bakx2025, bakx2025II} extended this work and compared this effective field theory (EFT) description to dark-matter–only simulations, demonstrating that the model can accurately describe the bispectrum within the statistical precision expected for a Stage-IV–like survey in the midly non-linear regime. More recently, \citet{gomes2026} presented predictions for the 3PCF of the aforementioned model and showed that higher-order contributions can lead to non-negligible differences as compared to simpler models for stage IV surveys. 

Furthermore, for galaxy-clustering, EFT methods have been applied to large hydrodynamic simulations \citep{ivanov2024, sullivan2025,shiferaw,zennaro2025}, which demonstrate that realistic galaxy-formation processes and baryonic feedback can be consistently absorbed into the EFT framework. 

In this work, we adopt a realistic setup in configuration space, employing statistics such as the 3PCF and the aperture mass statistic. Configuration-space statistics offer several observational advantages over Fourier space, most notably the straightforward treatment of the survey geometry. We measure third-order intrinsic alignments of galaxies (and of their host haloes) in the \textsc{FLAMINGO} hydrodynamical simulation suite \citep{Schaye2023, Kugel2023} and compare the results with predictions from the EFT of IA, as well as with several phenomenological models used in the literature. The \textsc{FLAMINGO} simulations are particularly well suited to this study: EFT is a perturbative framework whose validity is greatest on large, quasi-nonlinear scales, and testing it robustly therefore requires a simulation with sufficiently large volume. \textsc{FLAMINGO} provides this, with its large box size delivering the statistical power needed to probe these scales with high precision. We note that in this paper our focus is on the three-point signal, and we restrict our discussion of the two-point statistics to ensuring their consistency with the third-order results. A detailed study of the second-order alignment signal and its dependence on, e.g., baryonic feedback will be presented in \citet{herle26}.

This paper is organized as follows. In Section \ref{sec:sims} we discuss the FLAMINGO simulations used in this work, and how we extract galaxy and halo shapes from it. In Section \ref{sec:methods}, we go over the methods, describing the estimators, modelling and our inference methods. We show the results in Section \ref{sec:results}. Finally, we conclude in Section \ref{sec:concl}.

\section{FLAMINGO Simulations}\label{sec:sims}

The measurements in this work are obtained using the FLAMINGO suite of simulations \citep{Schaye2023, Kugel2023}, which were run with the SPHENIX smoothed particle hydrodynamics implementation \citep{borrow2022} in {\tt SWIFT} \citep{Schaller2024}. These simulations model neutrinos as massive particles following the method of \citet{elbers2021}, which effectively reduces shot noise. The simulations include radiative cooling and heating \citep{Ploeckinger2020}, star formation \citep{Schaye2008}, time-dependent stellar mass loss \citep{Wiersma2009}, and stellar feedback implemented through thermal and kinetic channels. The kinetic component follows the stochastic wind model of \citet{DallaVecchia2008} and uses the momentum-conserving pair-kicking scheme of \citet{Chaikin2023}, which conserves energy as well as linear and angular momentum. In the fiducial run, AGN feedback is implemented thermally \citep{Booth2009}. Halo identification is carried out with {\tt HBT-Herons} \citep{moreno2025}, an updated implementation of the {\tt HBT+} halo finder \citep{Han2018}. It uses a history-based tracking algorithm to consistently identify halos and subhalos across simulation snapshots.

In this work, we use the flagship run L2p8$\_$m9, which simulates a cubic comoving volume of $(2.8 \ \mathrm{Gpc})^3$ with $5040^3$ baryonic particles, an equal number of cold dark matter particles, and $2800^3$ neutrino particles. In this run, the average initial mass of the gas particles is $m_{\rm g} = 1.07 \times 10^9 \ {\rm M}_\odot$, and the cold dark matter particle mass is $m_{\rm cdm}= 5.65\times10^9 \ {\rm M}_\odot$. In addition to the fiducial 2.8 Gpc box, there are also smaller 1 Gpc boxes with varying implementations of baryonic physics. The impact of baryonic feedback on 2-pt. IA, using these simulation runs, is explored in \citet{herle26}.

\subsection{Galaxy and halo shapes}
Observed galaxy or halo shapes can be characterized as projected ellipses, described by an ellipticity  $\epsilon_0$ and a projected orientation angle $\phi$ with respect to a defined coordinate system. Alternatively, the shape can be expressed in terms of the complex number $\epsilon_c = \epsilon_{\rm 1}+ i \epsilon_2$ in a Cartesian basis, which relates to the orientation angle and ellipticity as follows:
\begin{equation}
    \epsilon_c = \epsilon_0 \exp( 2i \phi),
\end{equation}
where the ellipticity $\epsilon_0=|\epsilon_c|$. In this work, we derive object shapes using the simple, non-iterative inertia tensors provided by the FLAMINGO spherical overdensity aperture processor ({\tt SOAP}) catalogue \citep{2025soap}. Specifically, we adopt the unweighted (simple) inertia tensor, which is defined as a two-dimensional matrix:
\begin{equation}\label{eq:it}
I_{i j}=\frac{1}{M_{1/2}} \sum_{n=1}^N m^{(n)} x_i^{(n)} x_j^{(n)},
\end{equation}
where $M_{1/2} = \sum_i^N m_{i}$, is the sum over the mass of all bound stellar/dark matter particles within the half mass radius of the object, where we adopt the bound mass as our mass definition, and $x_i$ denotes the distance of particle $i$ from the halo center, which is taken to be the position of the most bound particle. Note that in our case $i$ and $j$ only run over the $x$ and $y$ coordinates, as we project our shapes over the $z$ direction. One could also opt for the reduced inertia tensor, in which case the particles are downweighted with a factor $1/r^2$, or use an iterative approach by repeatedly re-estimating the tensor using the inferred ellipsoidal geometry. These choices can matter, as they set the alignment amplitude ($A_{\rm IA}$) of the signal \citep{Kurita:2020hap, maion2025,herle26} and affect the signal-to-noise ratio (SNR), since the outskirts of galaxies and haloes tend to be more aligned than the inner part \citep{singh2016}. However, apart from a general amplitude scaling, higher-order bias parameters measured in the perturbative regime are found to be robust to the choice of inertia tensor, we confirm this in Appendix \ref{app:inertia_tensor}, consistent with other work \citep{akitsu2023, herle26}. Practically, reduced and iterative inertia tensors upweight the inner parts of haloes or galaxies. Furthermore, iterative definitions can in principle reduce shape noise. However, since these choices primarily rescale the overall alignment amplitude and do not qualitatively affect the results presented here, we adopt the simple, non-iterative inertia tensor.

The ellipticity \( \epsilon_0 \) and projected orientation angle \( \phi \) can be determined by calculating the eigenvalues \(\lambda_i\) and eigenvectors \(\mathbf{v}_i\) of the 2D inertia tensor. The ellipticity is given by  $\epsilon_0 = ({1-q})/({1+q})$,
where $q$ represents the sphericity, defined as 
$q = \sqrt{{\lambda_2}/{\lambda_1}}$ (see, e.g., \citealt{lammanhb2024}).
Here, \(\lambda_1\) and \(\lambda_2\) are the largest and smallest eigenvalues, respectively. The projected orientation angle \(\phi\) is defined as the angle between the first eigenvector (corresponding to the largest eigenvalue) and the reference axis.

\begin{figure}
    \centering
    \includegraphics[width=\linewidth]{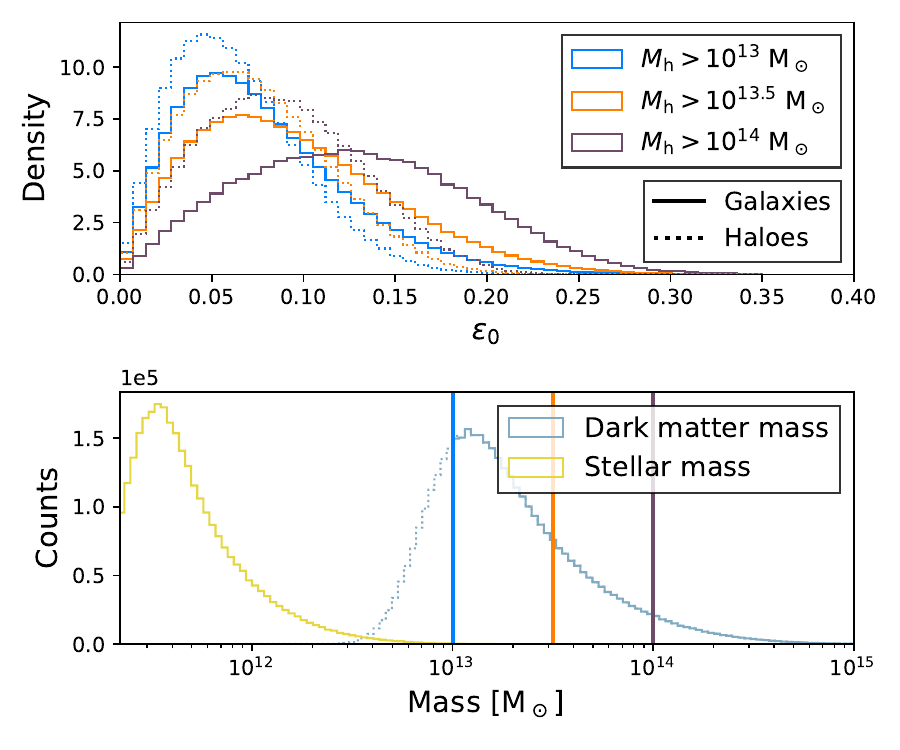}
    \caption{The upper panel shows the ellipticity distribution of either galaxies or haloes given a certain mass cut at $z=0$. The ellipticities are obtained via the simple, non-iterative, inertia tensor using only particles within the half mass radius of the object. The lower panel shows the corresponding mass ranges. A minimum of 300 stellar particles and a dark matter halo mass cut of $M_{\rm h} \geq 10^{13}\,{\rm M}_\odot$ are applied. The dotted histogram indicates dark matter haloes below this mass threshold that still satisfy the stellar particle cut; this subsample is not used in the analysis.}
    \label{fig:histograms}
\end{figure}

In our analysis, we use the FLAMINGO 2.8 Gpc box. We then select galaxies (and their haloes) based on several mass and particle cuts. First, we require that all galaxies contain at least 300 stellar particles to ensure that their shapes are well sampled \citep{tenneti2014, chisari2015, maion2025, herle2025}. Additionally, we make a dark matter cut of at least $10^{13} \ {\rm M}_\odot$ to ensure that we have a reasonably complete sample above this halo mass threshold. Throughout this work, when referring to haloes, we specifically mean dark matter haloes; accordingly, all halo inertia tensors are computed using only dark matter particles, and halo mass refers exclusively to the dark matter mass. We further restrict our sample to central galaxies and their haloes. After applying our other selection criteria, satellites constitute only about 10 percent of the sample, so their overall impact on the analysis is expected to be small. Nevertheless, we exclude them because they are likely to dilute the large-scale alignment signal, aside from any possible enhancement driven by the higher galaxy bias of the satellites and the corresponding increase in density weighting. Furthermore, their inclusion would complicate the mass dependence of the signal through variations in the satellite fraction. Applying these selection criteria yields approximately three million galaxies and haloes in our full sample. We show the ellipticity and mass distributions of our sample in Figure \ref{fig:histograms}. 

\section{Methods}\label{sec:methods}

\subsection{Statistical measures}
\subsubsection{2PCF and 3PCF}

Two-point correlations of galaxy or halo shapes and positions are quantified by the two-point correlation function (2PCF). Shapes are projected onto tangential ($\epsilon_{\rm t}$) and cross ($\epsilon_\times$) components:
\begin{equation}
    \epsilon\left(\mathbf{r}_\perp;\zeta\right) = \epsilon_{\rm t}\left(\mathbf{r}_\perp;\zeta\right) + i\epsilon_\times\left(\mathbf{r}_\perp;\zeta\right) = -\epsilon_c\left(\mathbf{r}_\perp\right) e^{-2i\zeta},
\end{equation}
where $\zeta$ denotes the projection angle, given by the polar angle of the pair separation in the two-point case. Furthermore, we have adopted the convention that $\epsilon_{\rm t} > 0$ corresponds to tangential alignments. For position–shape correlations, one then measures $\xi_{n+} = \langle \epsilon_{\rm t} \rangle$, where the cross component vanishes by parity. For shape–shape correlations, the combinations $\xi_{\pm} = \langle \epsilon_{\rm t}\epsilon_{\rm t} \rangle \pm \langle \epsilon_\times\epsilon_\times \rangle$ are usually measured. 

\begin{figure}
    \centering
    \begin{tikzpicture}[scale=4, every node/.style={font=\normalsize}]
      \coordinate (X1) at (0.5,0);
      \coordinate (X2) at (0,1);
      \coordinate (X3) at (1,1);
    
      \draw[ ->] (X1) -- (X2) node[pos=0.5, left=1pt] {$d_2$};
      \draw[ ->] (X1) -- (X3) node[pos=0.5, right=2pt] {$d_1$}; 
    \draw[dashed, -, red] (X1) -- (0.5,1.1) node[pos=0.5, right=2pt] {};

    \draw[fill = white, rotate = 10] (X1) ellipse (0.05cm and 0.1cm);
    \draw[fill = white, rotate = 40] (X2) ellipse (0.05cm and 0.1cm);
    \draw[fill = white, rotate = 80] (X3) ellipse (0.05cm and 0.1cm);
    \fill (X1) circle (0.2pt); 
    \fill (X2) circle (0.2pt); 
    \fill (X3) circle (0.2pt); 
    \draw (0.55,0) node[below right] {$r_1$};
    \draw (-0.05,1) node[left] {$r_3$};
    \draw (1,0.9) node[right] {$r_2$};
    
    \draw[-, thick] (0.4,0.2) to[out=45,in=135] (0.6,0.2);
    \draw[draw=white, fill=white] (0.5,0.35) node[fill=white, draw, text centered] {$\phi$};
    \end{tikzpicture}
    \caption{We define the triangles using the side-angle-side (SAS) congruence criterion. In the $\times$ projection, shapes located at vertex $r_1$ are projected along the red dashed line, shapes at $r_2$ along $d_1$ and shapes at $r_3$ along $d_2$. The red dashed line is oriented at an angle $\frac{1}{2}(\varphi_1 + \varphi_2)$ relative to the $x$-axis, where $\varphi_1$ and $\varphi_2$ denote the polar angles (measured from the $x$-axis) of $d_1$ and $d_2$ respectively.}
    \label{fig:triangles}
\end{figure}
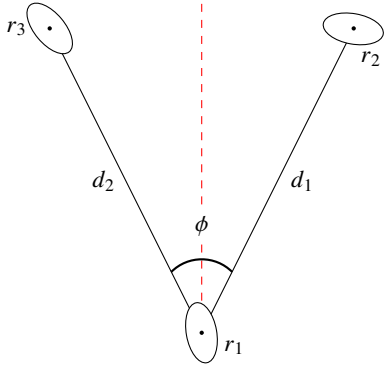

For three-point statistics, the situation is more complicated, as there is no unique projection direction associated with a triangle configuration; instead there are several choices that could be made. Given some projection $\zeta$, the `natural components' introduced by \citet{Schneider2003} are usually measured \citep{Porth2024,Burger2024, Secco2022, Fu2014}. In case of three shape fields, there are four complex valued natural components that do not mix and transform only by some phase factor under rotations. These are defined as follows:
\begin{equation}
\begin{aligned}
& 
\Gamma^{(0)}(d_1, d_2, \phi) = \left\langle\epsilon\left(\mathbf{r}_\perp^{1};\zeta_1\right) \epsilon\left(\mathbf{r}_\perp^{2};\zeta_2\right) \epsilon\left(\mathbf{r}_\perp^{3};\zeta_3\right)\right\rangle, \\
& 
\Gamma^{(1)}(d_1, d_2, \phi) =\left\langle\epsilon^*\left(\mathbf{r}_\perp^{1};\zeta_1\right) \epsilon\left(\mathbf{r}_\perp^{2};\zeta_2\right) \epsilon\left(\mathbf{r}_\perp^{3};\zeta_3\right)\right\rangle, \\
& 
\Gamma^{(2)}(d_1, d_2, \phi) =\left\langle\epsilon\left(\mathbf{r}_\perp^{1};\zeta_1\right) \epsilon^*\left(\mathbf{r}_\perp^{2};\zeta_2\right) \epsilon\left(\mathbf{r}_\perp^{3};\zeta_3\right)\right\rangle, \\
& 
\Gamma^{(3)}(d_1, d_2, \phi) = \left\langle\epsilon\left(\mathbf{r}_\perp^{1};\zeta_1\right) \epsilon\left(\mathbf{r}_\perp^{2}; \zeta_2\right) \epsilon^*\left(\mathbf{r}_\perp^{3};\zeta_3\right)\right\rangle.
\end{aligned}
\end{equation}
Here, $\mathbf{r}^i_\perp$ denotes the specific vertex of the triangle, and $d_1$, $d_2$ and $\phi$ the sides and opening angle, as shown in Figure \ref{fig:triangles}. In the case of galaxy-shape-shape (gII) cross-correlations, we interchange the shape field at $\mathbf{r}_\perp^{1}$ with a position field. In this case there are only two independent complex valued natural components. For the galaxy-galaxy-shape (ggI) correlation we interchange the shape fields at $\mathbf{r}_\perp^{2}$ and $\mathbf{r}_\perp^{3}$ with position fields. This results in there being only one independent complex-valued natural component. For these cross-correlations, the resulting correlation functions can be directly related to the 3PCFs defined by \citet{Schneider2005G3L}.

In this work, we measure the 3PCFs using the $\times$-projection introduced by \citet{Porth2024} and illustrated in Figure \ref{fig:triangles}. This projection is defined as
\begin{equation}
\zeta_1^{\times}=\frac{1}{2}\left(\varphi_1+\varphi_2\right), \quad
\zeta_2^{\times}=\varphi_1, \quad
\zeta_3^{\times}=\varphi_2,
\end{equation}
where $\varphi_1$ and $\varphi_2$ denote the polar angles of $d_1$ and $d_2$ respectively. The $\times$-projection is naturally aligned with the multipole basis employed in our analysis. However, in this projection, one galaxy is given a special status. Although this asymmetry is appropriate for cross-correlations, it is not ideal for autocorrelations. To address this, we adopt the centroid projection for the autocorrelation case, following \citet{Schneider2003}. The transformation from the $\times$-projection to the centroid projection is straightforward and can be accomplished by multiplying the 3PCF by a specific phase factor \citep{Porth2024}.

For cross-correlations, these correlation functions also contain disconnected contributions, as already noted by \citet{Schneider2005G3L}. This arises because the galaxy position field is probed as $N(\mathbf{r}_\perp) = \bar{N}(1+f(\delta\left(\mathbf{r}_\perp)\right))$. Factoring the correlators then introduces a contribution from the 2PCF due to multiplications involving the constant mean density $\bar{N}$. However, this issue can be addressed straightforwardly. For the 3PCF, we remove disconnected contributions by subtracting correlations involving random catalogues, as discussed in more detail in Section \ref{section:measurements}.

\subsubsection{Aperture statistics}
The $n$-point correlation functions in configuration space depend on a chosen projection axis and are therefore typically expressed as linear combinations of tangential and cross ellipticity components. For many analyses, it is convenient to reframe these components in terms of E- and B-modes \citep{crittenden2002}. The E-mode (B-mode), denoted $\epsilon_E$ ($\epsilon_B$), corresponds to the curl-free (gradient-free) part of the shape field and is most naturally defined in Fourier space. There, the complex shape $\epsilon_{\rm 1} + i\epsilon_2$ can be related to the E- and B-modes by the relation:
\begin{equation}
\left(\epsilon_{\rm 1}(\textbf{k}_\perp) + i\epsilon_2(\textbf{k}_\perp)\right)e^{-2i\phi}=  \epsilon_E(\textbf{k}_\perp) + i\epsilon_B(\textbf{k}_\perp),
\end{equation}
where $\phi$ is the polar angle of the wave vector $\mathbf{k}$ in the plane perpendicular to the line of sight (i.e., in the $k_x$–$k_y$ plane), with $k_z$ defining the line-of-sight direction.\footnote{We note that our $\epsilon_E$ is also called the convergence $\kappa_E$ in the context of weak lensing, as for example defined in \citet{Schneider2002}.} 

The 3PCFs, as defined in the previous section, contain all the third order information about the underlying shape field. However, they are typically noisy and difficult to interpret. For this reason, it is useful to compress them using the aperture mass statistic \citep{Schneider1998}. This approach offers two main advantages. First, it separates the signal into E- and B-mode components, which facilitates comparison with theoretical models. Second, it reduces the relatively large data vector to a more manageable size in an efficient manner, as most of the signal resides in the E-modes. 

The aperture mass is defined as follows
\begin{align}
    M_{\rm ap/\times}(R) &= \int \operatorname{d}^2 \textbf{r}_\perp Q_R(r_\perp) \epsilon_{t/\times}(\textbf{r}_\perp) \\
    &= \int \operatorname{d}^2 \textbf{r}_\perp U_R(r_\perp) \epsilon_{E/B}(\textbf{r}_\perp).
\end{align}
Here, $\epsilon_{t/\times}$ is taken relative to the centre of the aperture, $\epsilon_{E/B}(\textbf{r}_\perp)$ is defined by taking the inverse Fourier transform of $\epsilon_{E/B}(\textbf{k}_\perp)$, and $U_R$ is a compensated filter related to $Q_R$ through
\begin{equation}
    Q_R(r_\perp) = \frac{2}{r_\perp^2}\int_0^{r_\perp} \mathrm{d}x \ \ x \ U_R(x) - U_R(r_\perp).
\end{equation}
Note that, similar to the ellipticities, $M_{\rm ap}$ and $M_\times$ can be combined into one complex number $M = M_{\rm ap} + i M_\times$. Throughout this paper, we assume the kernel function introduced by \citet{crittenden2002}:
\begin{equation}
    U_R(r_\perp) = \frac{1}{2\pi R^2}\left(1 - \frac{r^2_\perp}{2R^2}\right) e^{-\frac{r_\perp^2}{2R^2}}.
\end{equation}
Similarly to the shapes, one can define the count aperture as
\begin{equation}
    N(R) = \int \operatorname{d}^2 \textbf{r}_\perp U_R(r_\perp) n(\textbf{r}_\perp),
\end{equation}
where $n$ is the number density of the galaxy/matter field. To extract information, we then take moments of these aperture mass quantities. For example, for $ \left\langle N(R_1) N(R_2) M_{\rm ap} (R_3)\right\rangle$,
\begin{multline}\label{eq:aperturedef}
    \left\langle N(R_1) N(R_2) M_{\rm ap} (R_3)\right\rangle = \\ \prod_i^3\int \mathrm{d}^2 \textbf{r}^{(i)}_\perp U_{R_i}\left({r}^{(i)}_\perp\right) \left\langle n\left(\boldsymbol{r}^{(1)}_\perp\right) n\left(\boldsymbol{r}^{(2)}_\perp\right) \epsilon_E\left(\boldsymbol{r}^{(3)}_\perp\right)\right\rangle.
\end{multline}
With similar expressions for the other third-order aperture mass statistics, corresponding to position-shape-shape (e.g. $\langle N M_{\rm ap}^2\rangle$) and shape-shape-shape (e.g. 
$\langle M_{\rm ap}^3\rangle$) correlations. Estimating aperture quantities directly from data is challenging. In particular, these estimators are sensitive to masking and gaps in the data: because the aperture mass is defined as a weighted integral over all shear values within a given radius, any missing region within that aperture distorts the integral and biases the resulting statistic. For this reason, the standard approach is to express the aperture mass in terms of the 3PCF. This transformation depends on the spin of the tracer field.

For the 3PCF the conversion functions were derived by \citet{Jarvis2004, Schneider2005, Schneider2005G3L}. These functions can then also be written as
\begin{equation}\label{eq:conversion}
    \mathrm{3AM}^j = \sum_i\int \mathrm{d}r_1 \mathrm{d}r_2 \mathrm{d}\phi \ A^j_i(r_1,r_2,\phi) \  \Gamma^j_i(r_1, r_2, \phi),
\end{equation}
where $i$ runs over the different natural components and we defined 3AM$^j$ = $\{\langle NNM \rangle, \langle NMM \rangle, \langle NMM^* \rangle, ...\}$, where the index $j$ selects the relevant aperture statistic. The specific kernels $A^j_i$ can be found in the cited papers. 

For three-point statistics, the aperture mass statistics have the additional benefit of selecting only the connected part of the 3PCF 
\citep{Schneider2005G3L}, a consequence of their compensated nature. Although we have already accounted for the disconnected contributions arising from the non zero average of the galaxy positions field at the level of the correlation functions themselves by subtracting the random contribution, computing the aperture mass of the disconnected part provides a useful consistency check, as it should vanish by construction.

Furthermore, we note that for the conversion from the 3PCF to the aperture mass to be exact, the integral in Eq. \ref{eq:conversion} must be evaluated over the full range of projected separations, from zero to infinity. In practice, however, this is impossible because the 3PCF can only be measured over a finite interval of scales. As a consequence, the E– and B–mode separation provided by the aperture mass statistics is only approximate: truncating the integration range leads to leakage between the modes \citep{kilbinger2006, shi2014}. In particular, power from $M_{\rm ap}$ is partially transferred to $M_\times$ due to the finite limits of the integral. Hence, the minimum (maximum) scale probed by the 3PCF should be sufficiently smaller (larger) than the aperture radius $R$. Alternatively, the aperture statistics could be modelled in the same way as the measurements themselves, by directly predicting the 3PCF and converting it using Eq. \ref{eq:conversion}. For instance, \citet{sugiyama2024} demonstrate that sufficiently fast predictions of third-order lensing statistics can be achieved with this approach. In this framework, the aperture cut-off scales are incorporated automatically, as well as any discrete binning effects that can become important. We do not explore this method further in this work.

\subsubsection{Estimators}\label{section:measurements}
We measure the following estimator for the 2PCF \citep{mandelbaum2011}, 
\begin{equation}\label{eq:2pcf}
    w_{nI}(r_\perp, r_\parallel \leq \Pi) =  \frac{\tilde{D}S}{RR},
\end{equation}
where we have defined $\tilde{D} = D -R$, to have a \citet{landy1993} like estimator. Specifically, $DS$ is defined as
\begin{equation}
    DS = - \sum_{\rm unique \ pairs} \epsilon_c \ e^{-2i  \phi},
\end{equation}
where $\epsilon_c$ is the ellipticity in a Cartesian frame and $\phi$ is the polar angle of the projected separation vector. $RR$ is the number of random galaxy pairs in this bin, rescaled to have the same amount of galaxies as our sample.

We consider only projected correlations along the line of sight: for each $r_\perp$ bin, we consider pairs that are at most a distance $\Pi$ away from each other, which we keep at a constant $20$ Mpc. In our analysis we do not model or include redshift-space distortions, since we work entirely in configuration space. We note, however, that in observational applications a sufficiently large projection length is important for suppressing redshift space distortion contamination \citep{lamman2025}.

We then extend these for the 3PCF as follows
\begin{equation}\label{eq:3pcfestimator}
    \Gamma^{\rm III}_0 = \frac{SSS}{RRR},  \ \ \    \Gamma^{\rm gII}_0 = \frac{\tilde{D}SS}{RRR},    \ \ \   \Gamma^{\rm ggI}_0 = \frac{S\tilde{D}\tilde{D}}{RRR},
\end{equation}
and similarly for the other natural components. Here, we impose a line of sight restriction that the second and third tracers need to be within a parallel distance of $\Pi$ to the first one, where we keep $\Pi$ fixed at 20 Mpc again. In the case of $\tilde{D}SS$, this means that the two shape-tracers need to be within a parallel distance of $\Pi$ of the position tracer. More precisely, the estimators are given by (largely following the notation of \citealt{Porth2024}) 
\begin{multline}
SSS = \sum_{i,j,k=1}^{N_{\mathrm{gal}}}
\left( -\epsilon_{\mathrm{c}, i}\,
{e}^{-\mathrm{i} (\varphi_{ij} + \varphi_{ik})} \right)
\left(- \epsilon_{\mathrm{c}, j}\,
{e}^{-2 \mathrm{i}\varphi_{ij}} \right) \\
\times
\left( -\epsilon_{\mathrm{c}, k}\,
{e}^{-2 \mathrm{i}\varphi_{ik}} \right)
\, \mathcal{B}_{ijk},
\end{multline}
where we kept the projection angles for every ellipticity explicit. Here, $\varphi_{ij}$ is the polar angle of $r_\perp^{(ij)}$, which is the projected distance between points $i$ and $j$. Furthermore,
\begin{multline}
\mathcal{B}_{ijk} \equiv \mathcal{B}\left(r_\perp^{(i j)} \in r^{(1)}_{\rm bin}\right) \mathcal{B}\left(r_\perp^{(i k)} \in r^{(2)}_{\rm bin}\right) \mathcal{B}\left(\phi_{i j k} \in \phi_{\rm bin}\right) \\ \times \mathcal{B}\left(r_\parallel^{(ij)}  < \Pi)\right)\mathcal{B}\left(r_\parallel^{(ik)}  < \Pi)\right).
\end{multline}
Here, $\mathcal{B}$ is only 1 if the condition is true, and $\phi_{ijk}$ is the opening angle of the triangle. To get the other natural components, one conjugates the relevant ellipticities and their phase. For cross-correlations, one replaces an ellipticity (and its corresponding complex angle and minus sign) by unity. In the position–position–shape case, the ellipticities at points $j$ and $k$ are set to unity, while in the position–shape–shape case, the ellipticity at point $i$ is set to unity.

Furthermore, we note that for the two-point functions, cross-correlations involving randoms and shapes ($RS$) can be safely neglected in our setup. In contrast, for the three-point statistics, the corresponding terms (e.g., $RDS$ or $RSS$) probe the disconnected contributions of the remaining two fields (see, e.g., \citet{philcox2021}). These contributions are non-negligible and must be subtracted to obtain an estimator of the connected component of the correlation function.\footnote{As noted above, this does not apply to the aperture mass statistics, for which such contributions vanish due to the compensated nature of the filter.}

It should come as no surprise that a na\"ive implementation of the estimator in Eq. \ref{eq:3pcfestimator} quickly becomes computationally infeasible, as the number of triplets scales as $\mathcal{O}(N^3)$. Although tree-based methods can significantly reduce computational cost, leading to more practical runtimes, they still demand substantial computational resources. A more efficient approach involves the use of a multipole decomposition. This method allows for a more efficient estimation by exploiting the underlying symmetries of the correlation functions, allowing it to be effectively treated as a two-point correlation function \citep{chen2005, slepian2015, philcox2022}. Although these methods were initially developed for clustering applications, recent work has generalised these to correlations including spin-2 tracers
\citep{Porth2024}, making them suitable for our work. 

The basic idea is to apply a Fourier transform to convert the $\phi$ binning into $n$-space via
\begin{equation} \Upsilon^i(d_1, d_2, \phi) = \sum_{n=-\infty}^{\infty} Z^i_n(d_1, d_2) e^{i n \phi}, 
\end{equation} 
where $\Upsilon^i = \{SSS, DSS, DDS, DDD, ...\}$. As demonstrated in the literature, this transformation significantly improves computational efficiency, as many of the resulting sums factorize. While the sum over $n$ is formally infinite, in practice it is sufficient to truncate the series at some finite $n_{\rm max}$.

To estimate the 3PCFs in the multipole basis, we use the well established {\tt TreeCorr} code \citep{Jarvis2004}.\footnote{\hyperref[]{https://github.com/rmjarvis/TreeCorr}} We then convert these into the angular 3PCF using the {\tt Orpheus} package \citep{porth2025}.\footnote{\hyperref[]{https://github.com/lporth93/orpheus}} For the aperture mass statistics, we consider the same estimators, making sure they are binned relatively fine, and convert them into aperture mass statistics using Eq. \ref{eq:conversion}. For this last step we use {\tt Orpheus} again. To compute the aperture mass statistics, we evaluate the 3PCF over radial scales ranging from $0.41$ to $560\,\mathrm{Mpc}$, using logarithmic bins with a width of $0.1$, and over $200$ angular bins, ranging from 0 to $2\pi$, using $n_{\rm max}=100$. This results in a total grid of $73^2 \times 200$ bins. In our context, the choice of the minimum scale, $r_{\rm min}$, is particularly important, as it must be sufficiently smaller than the aperture radius to ensure accurate and reliable measurements. Furthermore, for the aperture statistics we use a line-of-sight–integrated normalization, corresponding to an additional overall factor of $(2\Pi)^2$ relative to the 3PCF normalization.
\subsection{Modelling}\label{sec:modelling}
The shape `$\epsilon$' is assumed to comprise three distinct contributions:
\begin{equation}
    \epsilon = \epsilon_{\rm I} + \epsilon_{\rm G} + \epsilon_{\rm N},
\end{equation}
where $\epsilon_{\rm I}$ denotes the intrinsic alignment component, $\epsilon_{\rm G}$ is the result of gravitational lensing, and $\epsilon_{\rm N}$ represents the contribution of random noise. The latter can be reduced by increasing the sample size, while the second can be reduced by isolating close pairs along the line of sight.  In what follows, we focus on modelling the remaining term: intrinsic alignments.

We begin by introducing the 3D shape tensor \( g_{ij} \), which is a symmetric, traceless tensor, related to the 3D inertia tensor by:
\begin{equation}
    g_{ij}(\mathbf{x}) = \frac{\operatorname{TF}(I^{\rm 3D}_{ij})(\mathbf{x})}{\left\langle \operatorname{Tr}(I^{\rm 3D}_{ij}) \right\rangle},
\end{equation}
where \( \operatorname{TF} \) denotes the trace-free part of the tensor. As we only observe galaxy shapes in two dimensions, this tensor needs to be projected to a 2D tensor. We perform this projection on the sky using the trace free operator
\begin{align}
\mathcal{P}^{i j k l}(\hat{\mathbf{n}}) &=\operatorname{TF}\left(\mathcal{P}^{i k}(\hat{\mathbf{n}}) \mathcal{P}^{j l}(\hat{\mathbf{n}})\right), \\
\epsilon_{ij}(\mathbf{x}) &= \mathcal{P}^{ijkl}(\hat{\mathbf{{n}}})g_{kl}(\mathbf{x}),
\end{align}
which is built from the projection operator $\mathcal{P}_{ij} = \delta_{ij}-\hat{\textbf{n}}_i \hat{\textbf{n}}_j$, where $\hat{\textbf{n}}$ is the unit vector of the line of sight. The remaining two degrees of freedom can be expressed in terms of the familiar E- and B-mode basis using the projector operators $M^{E}_{ij}(\hat{\textbf{k}}, \hat{\textbf{n}})$ and $M^B_{ij}({\hat{\textbf{k}},\hat{\textbf{n}}})$
\begin{equation}\label{eq:los}
\begin{aligned}
\epsilon_E(\hat{\mathbf{k}}, \hat{\mathbf{n}}) & =\mathbf{M}_{i j}^E(\hat{\mathbf{k}}, \hat{\mathbf{n}}) g_{i j}(\mathbf{k}) ; \\
\epsilon_B(\hat{\mathbf{k}}, \hat{\mathbf{n}}) & =\mathbf{M}_{i j}^B(\hat{\mathbf{k}}, \hat{\mathbf{n}}) g_{i j}(\mathbf{k}).
\end{aligned}
\end{equation}
For their respective definitions and derivations we refer to \citet{bakx2025}. However, we note that if $g_{ij}$ is given in linear order in perturbation theory, the B-mode vanishes: $M^B_{ij}(\hat{\textbf{k}},\hat{\textbf{n}})g_{ij}(\textbf{k}) = 0$. In the most general setting, the tensor bispectrum $B_{ijklrs}$ is constructed from three tensors $S_{ij}$, each of which contains both a trace and a traceless component, with the trace part corresponding to scalar perturbations
\begin{equation}
S_{ij}(\mathbf{x})
= \frac{1}{3}\delta_{ij}\,\delta(\mathbf{x}) + g_{ij}(\mathbf{x}).
\end{equation}
For scalar tracers, such as the positions of dark matter particles, the
corresponding projection is then simply given by the Kronecker delta,
\(M^\delta_{ij}=\delta_{ij}\). These tensors can be combined to obtain the tensor bispectrum
\begin{equation}
\begin{aligned}
\left\langle S_{i j}\left(\mathbf{k}_1\right) S_{k l}\left(\mathbf{k}_2\right) S_{r s}\left(\mathbf{k}_3\right)\right\rangle & =(2 \pi)^3 \delta^{(D)}\left(\mathbf{k}_1+\mathbf{k}_2+\mathbf{k}_3\right) \\
& \times B_{i j k l r s}\left(\mathbf{k}_1, \mathbf{k}_2, \mathbf{k}_3\right),
\end{aligned}
\end{equation}
then, the bispectrum is obtained by applying the appropriate projection operators to combinations of $S_{ij}$:
\begin{equation}\label{eq:bispecprojection}
B_{XYZ}
= \mathbf{M}_{ij}^X(\hat{\mathbf{k}}_1,\hat{\textbf{n}})\,
  \mathbf{M}_{kl}^Y(\hat{\mathbf{k}}_2,\hat{\textbf{n}})\,
  \mathbf{M}_{rs}^Z(\hat{\mathbf{k}}_3,\hat{\textbf{n}})\,
  B_{ijklrs}.
\end{equation}
For $X,Y,Z\in \delta,E,B$. To illustrate, in the case of matter-matter-shape statistics, $B_{ijklrs}$ reduces to
\begin{equation}
 (2 \pi)^3\delta^{(D)}(\mathbf{k}_1+\mathbf{k}_2+\mathbf{k}_3)\,
B_{\delta \delta,ij}
=
\left\langle
\delta(\mathbf{k}_1)\,
\delta(\mathbf{k}_2)\,
g_{ij}(\mathbf{k}_3)
\right\rangle ,
\end{equation}
where we have already projected out the scalar indices. The E-mode, $B_{\delta \delta E}$, can then be constructed by applying $M^E_{ij}$ to this spectrum. The remaining bispectra follow analogously by interchanging the $\delta$'s with $g$'s in the appropriate positions.

These spectra can be related to the measured aperture mass statistics by evaluating the integrals derived in Appendix \ref{app:losproj}. Under the assumption that the bispectra depend only on the three wave vectors and on two angles with respect to the line of sight \citep{bakx2025}, the expression for $\langle N N M_{\rm ap}\rangle$ takes the form
\begin{multline}
\left\langle N(R_1) N(R_2) M_{\rm ap}(R_3) \right\rangle = \\
\begin{aligned}
&\frac{1}{(2\pi)^5} \int 
  k^{(1)}_\perp \, \mathrm{d}k^{(1)}_\perp \, \mathrm{d}k^{(1)}_\parallel \,
  k^{(2)}_\perp \, \mathrm{d}k^{(2)}_\perp \, \mathrm{d}k^{(2)}_\parallel \, \mathrm{d}\phi \\
&\quad \times \tilde{W}_\Pi(k^{(1)}_\parallel) \, \tilde{W}_\Pi(k^{(2)}_\parallel) \,
             \tilde{U}_{R_1}(k^{(1)}_\perp) \, \tilde{U}_{R_2}(k^{(2)}_\perp) \\
&\quad \times \tilde{U}_{R_3}(||\mathbf{k}^{(1)}_\perp + \mathbf{k}^{(2)}_\perp||) \,
             B_{\delta \delta E}(k^{(1)}_\perp, k^{(2)}_\perp, k^{(1)}_\parallel, k^{(2)}_\parallel, \phi),
\end{aligned}
\end{multline}
where $\tilde{W}_\Pi(x) = 2\sin(\Pi x)/x$ is the Fourier transform of the real-space top-hat window of half-width $\Pi$, which selects the matter distribution within a line-of-sight interval around the shape tracer. Note that the tracer in the third position, in this case the shape, thus plays a special role: the other tracers are projected relative to this one. The function $\tilde{U}_R(k)$ is the Fourier transform of the compensated aperture mass filter,
\begin{equation}
\tilde{U}_R(k) = \frac{k^2 R^2}{2}\,
                 \exp\!\left(-\frac{k^2 R^2}{2}\right).
\end{equation}
This filter peaks at $kR=\sqrt{2}$ and is relatively narrow in Fourier space. Consequently, aperture mass statistics predominantly probe (projected) triangle configurations in Fourier space with characteristic side lengths $k_i \sim \sqrt{2}/R_i$.

The expressions above apply to all other combinations of projected positions and shapes as well; only the bispectrum $B_{XYZ}$ and the ordering of the vertices need to be adjusted so that the appropriate triangle vertex is treated as the central one. For example, in position-shape-shape configurations we use $B_{EE\delta}$ in our convention, where the shape tracers are required to lie within a line-of-sight distance $\Pi$ of a central overdensity.

\subsubsection{EFT of IA}
We give a brief overview of the Eulerian EFT of IA, and refer the reader for a more detailed description to \citet{vlah2020, bakx2023, bakx2025}. We model the shape tensor $g_{ij}$. This tensor is then expressed in terms of the following operators,
\begin{equation}\label{eq:eft}
g_{i j} \mathbf{\ \approx \ } b_K K_{i j}+b_{\delta K} \delta K_{i j}+b_{K K} \mathrm{TF}\left(K^2\right)_{i j} + b_t t_{ij} + \ldots,
\end{equation}
where the operators are expressed in the basis also used by e.g. \citet{blazek2019,schmitz2018}, and we refer to \citet{bakx2023} for their exact definitions. The dots denote higher order terms not accounted for in this analysis. Furthermore, the EFT contains contributions from stochastic operators. However, these do not contribute in configuration space, and therefore we ignore them in our analysis. The normalization of the parameters is in principle arbitrary, and we use the same normalization convention as \citet{vlah2020,bakx2023,bakx2025II}. We note that in our convention $b_K$ can be connected to the more commonly used $A_{\rm IA}$ parameter via
\begin{equation}
    b_K = -2 A_{\mathrm{IA}} C_1 \frac{{\rho}_{\rm crit}\Omega_{\rm m,0}}{D(z)}.
\end{equation}
$\Omega_{\rm m,0}$ denotes the matter content at $z=0$ and $A_{\rm IA}$ represents the alignment amplitude, which parametrizes the response after conventionally fixing $C_1 \rho_{\rm crit} = 0.0134$. $D(z)$ is the growth function normalised to unity at redshift zero. Furthermore, following the aforementioned studies, we refer to the parameters as density weighting ($b_{\delta K}$), tidal torquing ($b_{KK}$) and velocity shear ($b_t$). Instead of the basis in Eq. \ref{eq:eft}, one could equivalently choose to work in the `EFT' basis $\{b^g_1, b^g_{21}, b^g_{22}, b^g_{23}\}$, employed by \citet{vlah2020, bakx2023, bakx2025}.\footnote{The two bases are related via 
$b_K =b_1^g, b_{\delta K} =\frac{1}{3}\left(2 b_{2,1}^{\mathrm{g}}+b_{2,3}^{\mathrm{g}}\right),  b_{KK}=b_{2,1}^{\mathrm{g}}-b_{2,3}^{\mathrm{g}}, b_t =\frac{7}{8}\left(b_{2,3}^{\mathrm{g}}-b_{2,2}^{\mathrm{g}}\right)$, see  \citet{vlah2020}.}

The specific bispectra are obtained from
\begin{multline}
B_{ijklrs}^{\alpha\beta\gamma,\mathrm{det,tree}}
(\mathbf{k}_1,\mathbf{k}_2,\mathbf{k}_3)
= 2\,
\mathcal{K}_{ij}^{\alpha,(1)}(\mathbf{k}_1)
\mathcal{K}_{kl}^{\beta,(1)}(\mathbf{k}_2)
\mathcal{K}_{rs}^{\gamma,(2)}(\mathbf{k}_1,\mathbf{k}_2) \\
\times\, P_{\rm L}(k_1)\, P_{\rm L}(k_2)
+ 2\ \mathrm{perm.},
\end{multline}
where $\alpha,\beta$ and $\gamma$ denote the tracer, the superscript number indicates that one of the three kernels is evaluated at second order and $P_{\rm L}$ is the linear power spectrum. This tensor bispectrum can be converted into e.g. $B_{\delta \delta E}$ using Eq. \ref{eq:bispecprojection}, and the kernel functions will contain the contributions of the different operators; we refer to Appendix A of \citet{bakx2025} for their exact definitions. 

At this stage, it is useful to compare the EFT and TATT models. The TATT model \citep{blazek2019}, short for Tidal Alignment and Tidal Torquing model, is in principle built from the same set of operators as the EFT. In practice, however, TATT implementations typically omit the velocity–shear operator ($b_t$) (see e.g. \citealt{samuroff2021,des2022, 2025fabian}), although we note that \citet{schmitz2018} included this term in their bispectrum analysis. Furthermore, for the power spectrum, the TATT model substitutes the fully non linear power spectrum in for the leading order term. To assess the importance of the velocity–shear contribution, we consider a version of the EFT in which this operator is explicitly set to zero. We refer to this model as EFT - no VS.

The EFT we will employ is formulated in \textit{Eulerian space}. However, it is also possible to work instead in \textit{Lagrangian space}, and subsequently advect the results to Eulerian space. This approach allows for the derivation of \textit{co-evolution relations}: expressions that relate higher-order parameters to the linear parameter. These relations are obtained under the assumption that alignment is linear in Lagrangian space, with the higher-order terms arising from the \textit{advection process} that maps quantities from Lagrangian to Eulerian space. This is known as the linear Lagrangian bias ansatz (LLB). Previous work has shown that this gives qualitatively good results for haloes \citep{akitsu2023, chen2024, maion2024}. The mapping between parameters is as follows:
\begin{align}
    b_{KK} &= -b_K, \\
    b_{\delta K} & = (b_1 - \frac{2}{3}) b_K, \\
    b_t &= \frac{5}{2}b_K.    
\end{align}
Here, $b_1$ is the linear galaxy bias, which becomes relevant for the density weighting term. We note that this is not expected to hold per se, any deviation merely implies a non-linear alignment in Lagrangian space. To test this ansatz, we will consider a variant of the EFT where we fix $b_t$ and $b_{KK}$ to the co-evolution relations and fit $b_K$ and $b_{\delta K}$. We will refer to this as EFT - LLB.

Hence, we will consider three models that contain non-linear shape bias: EFT, EFT - no VS and EFT - LLB, having 4, 3 and 2 free parameters. In our setup, EFT - no VS and EFT - LLB are subsets of the full EFT. However, they are not consistent with each other, as the EFT - no VS model does not satisfy the co-evolution relations.

\subsubsection{Non-linear alignment model (NLA)}
The linear alignment model \citep{catelan2001, hirata2004} posits that the shape distortions are proportional to the trace-free part of the tidal field of the gravitational potential, this can be understood as the first order term in a perturbative expansion like the one discussed in the previous section, and is therefore expected to work only at large scales. In this picture, we have
\begin{equation}
    g_{ij} = b_K K_{ij},
\end{equation}
which after applying the projection in Eq. \ref{eq:los} gives the more familiar
\begin{align}
    \epsilon_E &= \frac{b_K}{2}(1-\mu^2)\delta(\textbf{k},z), \\
    \epsilon_B &= 0,
\end{align}
 where $\delta$ is the density contrast and $\mu=\textbf{k} \cdot \hat{\textbf{n}}$. In the purely linear model, the intrinsic alignment bispectrum vanishes by Wick’s theorem. However, it is common to expand the matter overdensity to all orders by using non-linear matter spectra fitted from simulations. This prescription is referred to as the \emph{non-linear alignment} (NLA) model \citep{bridle2007}. In this case, the predicted bispectrum inherits all non-linear structure from the non-linear dark matter field, while higher-order IA terms are effectively neglected. In other words, this expansion takes the form of Eq. \ref{eq:eft} with the higher order parameters set to 0 and the matter field expanded to all orders.  In our implementation, we use the BiHalofit implementation for the non-linear bispectrum \citep{takahashi2020}:
 \begin{equation}
     B^{\rm NLA}_{\delta \delta E}(k_1,k_2,k_3,\mu_3)=\frac{b_K}{2}(1-\mu_3^2)B^{\rm bhf}_{\delta \delta \delta}(k_1,k_2,k_3),
 \end{equation}
where $\mu_3$ denotes $\hat{\textbf{k}}_3\cdot\hat{\textbf{n}}$. Strictly speaking, this model is not self-consistent, since the higher-order alignment bias terms enter at the same perturbative order as the non-linear matter contributions. Nevertheless, for the BOSS LOWZ sample, \citet{linke2024} report that an NLA-inspired prescription suffices,\footnote{We note, however, that our estimator for the 3PCF is normalized relative to a random sample, which is different from what was done in \citet{linke2024}.} and it has been the standard choice for third-order shear analyses \citep{Burger2024, des3pt2025, sugiyama2025}.

\subsection{Likelihood analysis}
Using the theoretical framework and measurements described above, we perform a likelihood analysis to estimate the model parameters. We adopt a Gaussian likelihood function $L$, expressed in terms of the data vector $D$, the covariance matrix $C$, and the model prediction $M(\theta)$ for a set of parameters $\theta$:
\begin{equation}
-2 \ln L(\theta) =
\sum_{i=1}^{N_D} \sum_{j=1}^{N_D}
\bigl[D_i - M_i(\theta)\bigr]
C^{-1}_{ij}
\bigl[D_j - M_j(\theta)\bigr],
\end{equation}
where $N_D$ is the number of data points.

The covariance matrix is estimated using a jackknife procedure. 
We partition the simulation volume into \(3 \times 4^{3}\) subvolumes, each of size \(700^{2} \times 233.33\,\mathrm{Mpc}^{3}\). By omitting one subvolume at a time, we construct \(3 \times 4^{3}\) jackknife samples. The subvolumes are intentionally noncubic, as our analysis probes much larger scales in the transverse (sky-plane, \(x\)–\(y\)) directions than along the line of sight (\(z\)).

Since the covariance matrix is estimated using jackknife resampling, it has a limited effective resolution given by
\begin{equation}
    \Delta C_{ij} \simeq \sqrt{\frac{2}{N_{\rm jk}}},
\end{equation}
where $N_{\rm jk}$ is the number of jackknife regions. To regularize the covariance matrix, we follow the approach of \citet{gaztanaga2005}. First, we rescale the data vector and covariance matrix by their variances. We then perform a Singular Value Decomposition (SVD):
\begin{equation}
\hat{C}_{i j} = \left(U_{i k}\right)^{\dagger} D_{k l} V_{l j},
\end{equation}
where $D_{kl}$ is a diagonal matrix with elements $\lambda_{i}^{2}$. To enforce the resolution criterion, we retain only those modes satisfying
\begin{equation}
    \lambda_{i}^{2} > \sqrt{\frac{2}{N_{\rm jk}}}.
\end{equation}
This procedure removes noisy modes and yields a well-conditioned covariance matrix. An additional benefit is that it compresses the data vector to a size significantly smaller than the number of jackknife patches. Despite this compression, we still apply the Hartlap correction \citep{hartlap2007}. Strictly speaking, this correction is derived under the assumption of independent samples and is therefore not formally valid for jackknife estimates. However, \citet{favole2021} demonstrated that it nevertheless provides a reasonable correction for the inverse covariance matrix in this context. We therefore adopt the standard Hartlap factor, $
f_{\mathrm{H}}=({N_{\text {JK }}-N_{\text {bin, SVD }}-2})/({N_{\text {JK }}-1})$, where, following \citet{philcox2021mocks}, $N_{\text {bin, SVD }}$ denotes the number of bins in the compressed data vector.

The posteriors are then sampled using the \texttt{Nautilus} nested sampler \citep{nautilus}, and the resulting posterior distributions are processed with \texttt{GetDist} \citep{lewis2025}. Throughout this work, we use uninformative priors. Using the best fit parameters, we assess the goodness of fit using the reduced chi-squared:
\begin{equation}
    \chi^2_{\rm red} = \frac{\chi^2}{N_{\rm d.o.f.}},
\end{equation}
where $\chi^2$ is the chi-squared statistic using the compressed data vector and covariance, and $N_{\rm d.o.f.}$ is the number of degrees of freedom. Furthermore, we assess the significance of the total signal using the signal-to-noise ratio. We adopt the definition introduced by \citet{Secco2022},
\begin{equation}
\mathrm{SNR} \equiv
\begin{cases}
\sqrt{\chi^2 - N_{\text{bin,SVD}}} & \text{if } \chi^2 > N_{\text{bin,SVD}} + 1, \\
\text{null} & \text{otherwise},
\end{cases}
\end{equation}
where $N_{\text{bin,SVD}}$ is the number of elements in the compressed data vector. Here, $\chi^2$ denotes the chi-square statistic computed from the compressed data vector and covariance with respect to the null (no-signal) model. In the limit $\chi^2 \gg N_{\text{bin,SVD}}$, this expression reduces to the commonly used approximation $\mathrm{SNR} \simeq \sqrt{\chi^2}$.

\section{Results}\label{sec:results}

\begin{figure}
    \centering
    \includegraphics[width=0.8\linewidth]{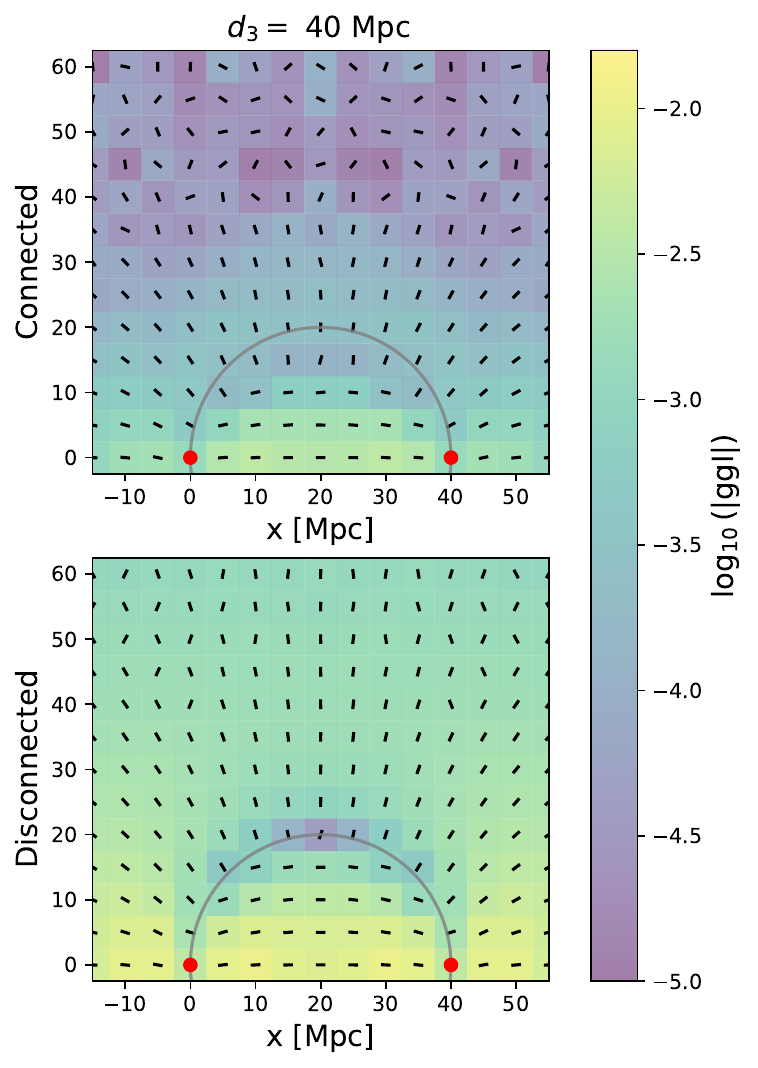}
    \caption{Alignment around galaxy pairs (red dots) separated by a distance $d_3$. The colour map represents the amplitude of the total alignment signal on a logarithmic scale. The black lines indicate the orientation of the galaxies. The grey circle has a radius of $d_3/2$ and is centered at $(d_3/2,\, 0)$; according to Thales' theorem, any triangle formed with points inside this circle has an opening angle larger than $90^\circ$, while triangles formed with points outside the circle have an opening angle smaller than $90^\circ$.}
    \label{fig:stacked_galaxy}
\end{figure}

\begin{figure}
    \includegraphics[width = \linewidth]{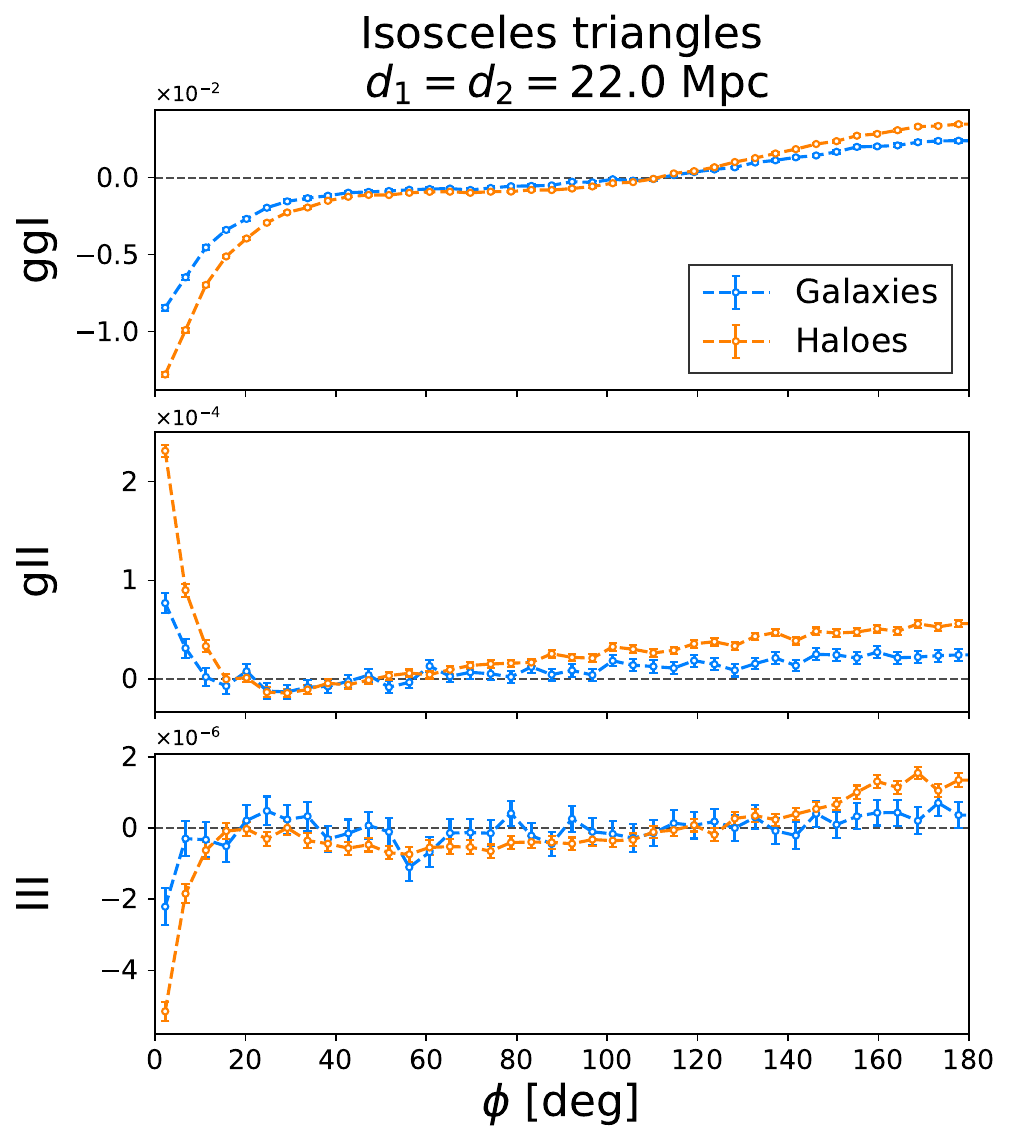}
    \caption{The $\phi$ dependence of the tangential components (obtained from the real part of the appropriate linear combinations of 3PCFs) of the connected 3PCF for position-position-shape correlations (ggI), position-shape-shape correlations (gII) and shape-shape-shape correlations (III).  Here we show our main sample of galaxies and haloes with $M_{\rm h} > 10^{13}$ M$_\odot$ at $z=0$. Note that $\epsilon_t>0$ means tangential alignment, and that these measurements are defined with respect to galaxy or halo positions, rather than to the dark matter particle positions considered later.}\label{fig:3pcf}
\end{figure}

\subsection{3PCF of IA}
In Figure \ref{fig:stacked_galaxy}, we illustrate the galaxy-galaxy-shape 3PCF, visualized as the pattern of galaxy shapes around two galaxies separated by a distance $d_3$. We show both the connected contribution (i.e., the full 3PCF with disconnected terms subtracted) and the disconnected 3PCF. The grey semi-circle is centred at the midpoint between the two galaxies, with a radius of $d_3/2$. By Thales' theorem, any point lying on this circle forms a right angle ($\phi = 90^\circ$) with respect to the two galaxies. As a result, all points within the circle correspond to triangles with opening angles $\phi > 90^\circ$, while points outside the circle correspond to  triangles with $\phi < 90^\circ$. If the connected contribution is zero, the expected alignment pattern is simple, as it follows directly from the two-point signal: points inside the circle align with the $x$-axis, while points outside the circle point towards a point on the line between the galaxies. At an opening angle of $\phi = 90^\circ$, thus all points on the grey circle, the shear of the two position galaxies interfere, strongly suppressing the signal. In the special case of an isosceles triangle with $\phi = 90^\circ$, where $d_1 = d_2$, this  interference leads to a complete cancellation of the disconnected contributions, resulting in no net alignment from these terms. This interference effect is clearly visible in the bottom panel of the figure, where only the disconnected contributions to the 3PCF are shown.

This differs in the connected case: the excess alignment due to non-linear structure formation. Here, we still see a similar turnover in the alignment signal, where it goes from pointing towards the connecting line, to aligning parallel with the connecting line. However, this turnover happens at a significantly larger opening angle compared to the disconnected case. If interpreting the line between the two galaxies as a filament \citep{clampitt2017,epps2017, xia2020}, this implies an excess alignment aimed towards filaments, consistent with observations by e.g. \citet{chen2019, laigle2025}. 

In Figure \ref{fig:3pcf}, we present the angular dependence of the 3PCF, illustrating all three combinations of shapes and positions. In all cases, the signal is stronger for haloes than for galaxies. Notably, we observe that for each combination, the signal peaks for more collinear triangle configurations: those in which the triangle's vertices lie approximately along a straight line. This feature has also been observed in the matter bispectrum (see, e.g., \citealt{scoccimarro1999}). Of the two nearly collinear options, most signal is in the squeezed case, where the angle is very small. For the galaxy-galaxy-shape correlation we see similar behaviour as in Figure \ref{fig:stacked_galaxy}, as we note a turnover point at $\phi > 90^\circ$.

\subsection{Third order aperture mass statistics and modelling}

\begin{figure}
    \centering
    \includegraphics[width=\linewidth]{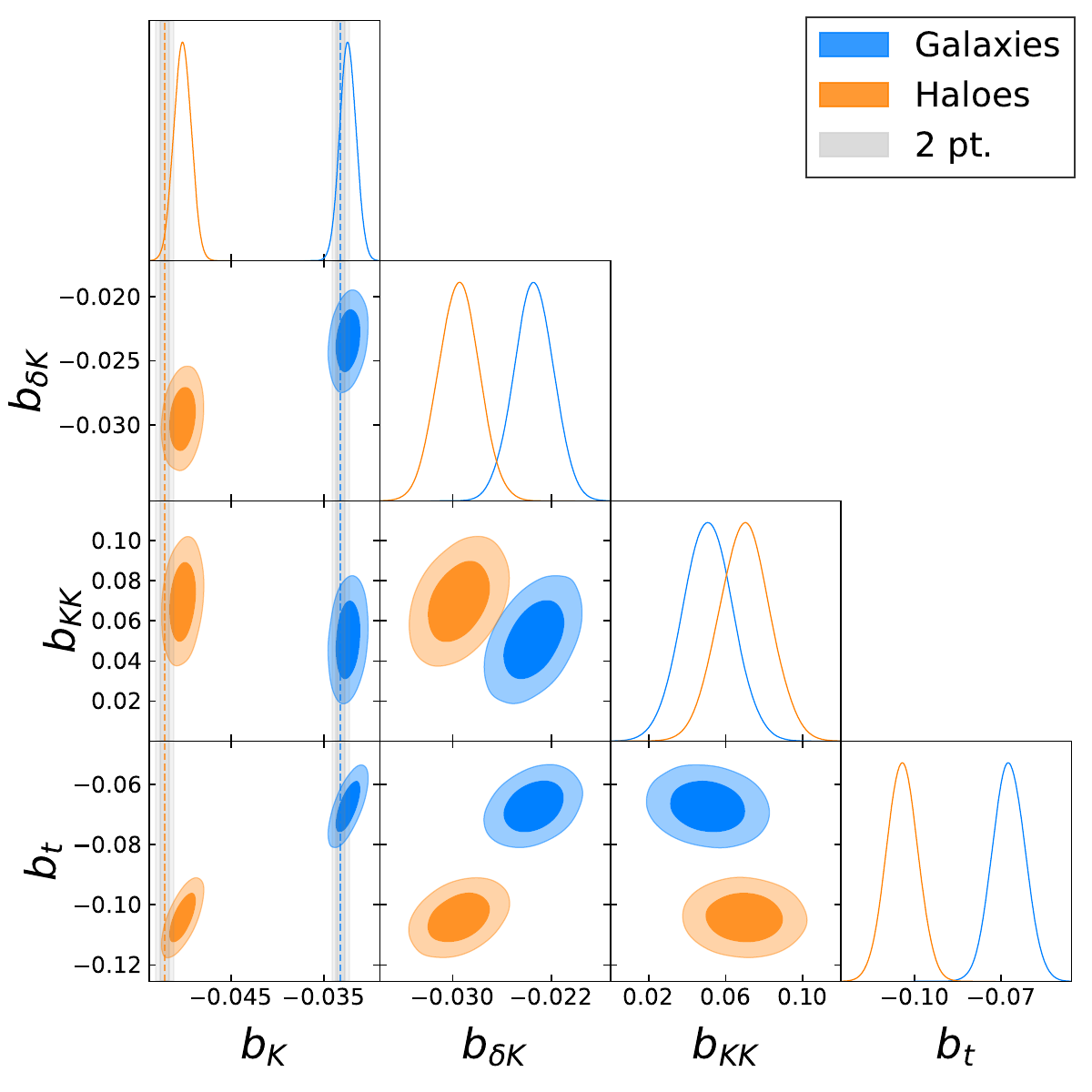}
    \caption{Posteriors of the EFT parameters fitted to  $\langle NNM_{\rm ap}\rangle$ of both galaxies and haloes. The grey lines show the alignment amplitude obtained from two-point statistics at scales $\geq$ 40 Mpc. The one (two) sigma around the two-point estimate is shown as a shaded region.  We show the posteriors for our main samples of $M_{\rm h} > 10^{13}$ M$_\odot$ at $z=0$ omitting configurations with radii $R_i$ smaller than 26 Mpc for the three-point statistics.}
    \label{fig:eft_posterior}
\end{figure}

\begin{figure}
    \centering
    \includegraphics[width=\linewidth]{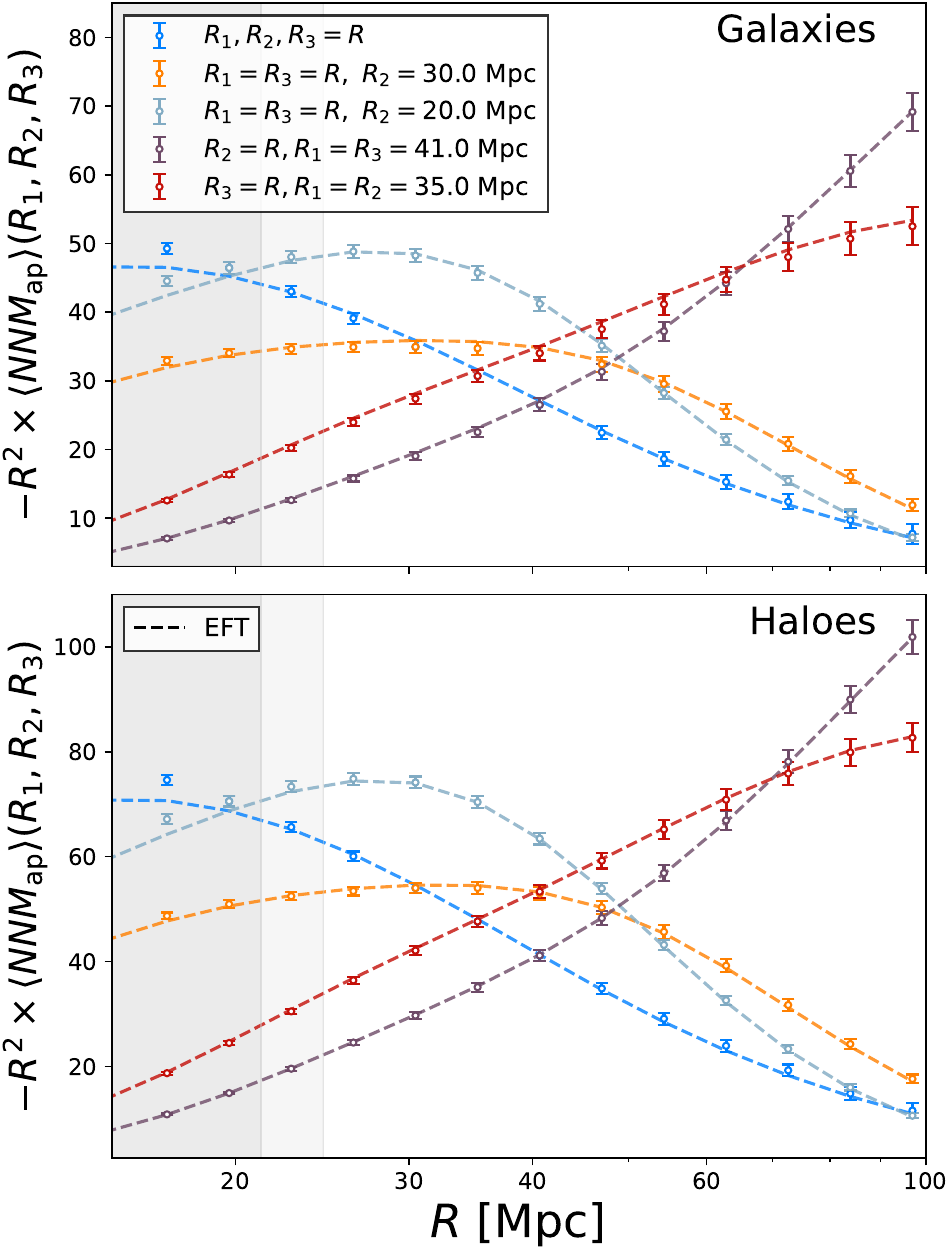}
    \caption{Measured $\langle NNM_{\rm ap}\rangle$ signal for several aperture radius configurations of the matter–matter–shape statistic. We show our main sample of $M_{\rm h} > 10^{13}$ M$_\odot$ at $z=0$. The corresponding theoretical predictions, using the best fit values from Eq. \ref{eq:bias_params}, are overplotted. Only a subset of the full data vector is shown to avoid overcrowding the figure. The leftmost grey-shaded region marks scales which, if included, would cause the reduced chi-square, $\chi^2_{\rm red}$, to deviate from unity by more than $2\sigma$, while the second grey-shaded region indicates the more conservative scale cut adopted for parameter constraints. Note that this visualization is only indicative: all data points with any $R_i$ below the scale cut to the right of the dashed lines are excluded from the fits.}
    \label{fig:data_fits}
\end{figure}

To connect these measurements with our models, we employ the aperture mass statistic. As discussed previously, the aperture mass statistic separates into E- and B-modes, making it a useful real-space analogue of the bispectrum. To model this statistic, we employ the formalism introduced in Section \ref{sec:modelling}. In order to disentangle the higher-order alignment bias from the higher-order galaxy bias, we correlate the galaxy shapes with randomly sampled dark matter particles. This setup is more analogous to that used in weak lensing surveys, where one directly probes matter–shape correlations rather than galaxy position-shape correlations. However, for a direct measurement of intrinsic alignments, obtained by correlating galaxy positions with galaxy shapes, it might be necessary to introduce additional parameters to account for higher-order galaxy biasing.

We fit the EFT model to the $\langle N N M_{\rm ap} \rangle$ statistic, using all aperture radius configurations with $R_i \geq 26\,\mathrm{Mpc}$ to constrain the model parameters.\footnote{This choice is conservative; as discussed later, the model begins to break down on smaller scales.  Using the breakdown criterion of \citet{bakx2023}, defined as a joint requirement on $\chi^2_{\rm red}$ and the Figure of Bias of the inferred $b_K$, we find that, relative to the 2-pt. $b_K$, the model remains valid down to scales of $\sim 22\,\mathrm{Mpc}$.  Nevertheless, we choose this scale cut to ensure our parameter constraints are fully converged.} We focus on $\langle N N M_{\rm ap} \rangle$ because it provides the highest signal-to-noise ratio among the three statistics. Nevertheless, we verify that the resulting model remains consistent with $\langle N M^2_{\rm ap} \rangle$ and $\langle M^3_{\rm ap} \rangle$. Furthermore, we discuss the B-mode statistic, $\langle NNM_\times \rangle$, in Appendix \ref{app:b-mode}. For our main galaxy sample, consisting of galaxies residing in haloes with ${ M}_h > 10^{13}\,{\rm M}_\odot$, the best-fit EFT parameter values obtained from our Markov Chain Monte Carlo analysis of the data vector are
\begin{equation}
\begin{aligned}
    b_K & = -0.0324 \ \pm \ 0.0009 \ \ \ (A_{\rm IA} = 3.96 \ \pm \ 0.11), \\
    b_{\delta K} &= -0.0234 \ \pm \ 0.0016, \\
    b_{KK} &= 0.051 \ \pm \ 0.0130, \\
    b_t &= -0.0675 \ \pm \ 0.0056.
\end{aligned}
\label{eq:bias_params}
\end{equation}
 The full 2D posteriors are shown in Figure \ref{fig:eft_posterior}. All parameters deviate from zero at several $\sigma$. We therefore find clear evidence for negative density-weighting ($b_{\delta K}$) and velocity-shear ($b_t$) contributions, as well as a positive tidal-torquing ($b_{KK}$) contribution.
\begin{figure}
    \centering
    \includegraphics[width=\linewidth]{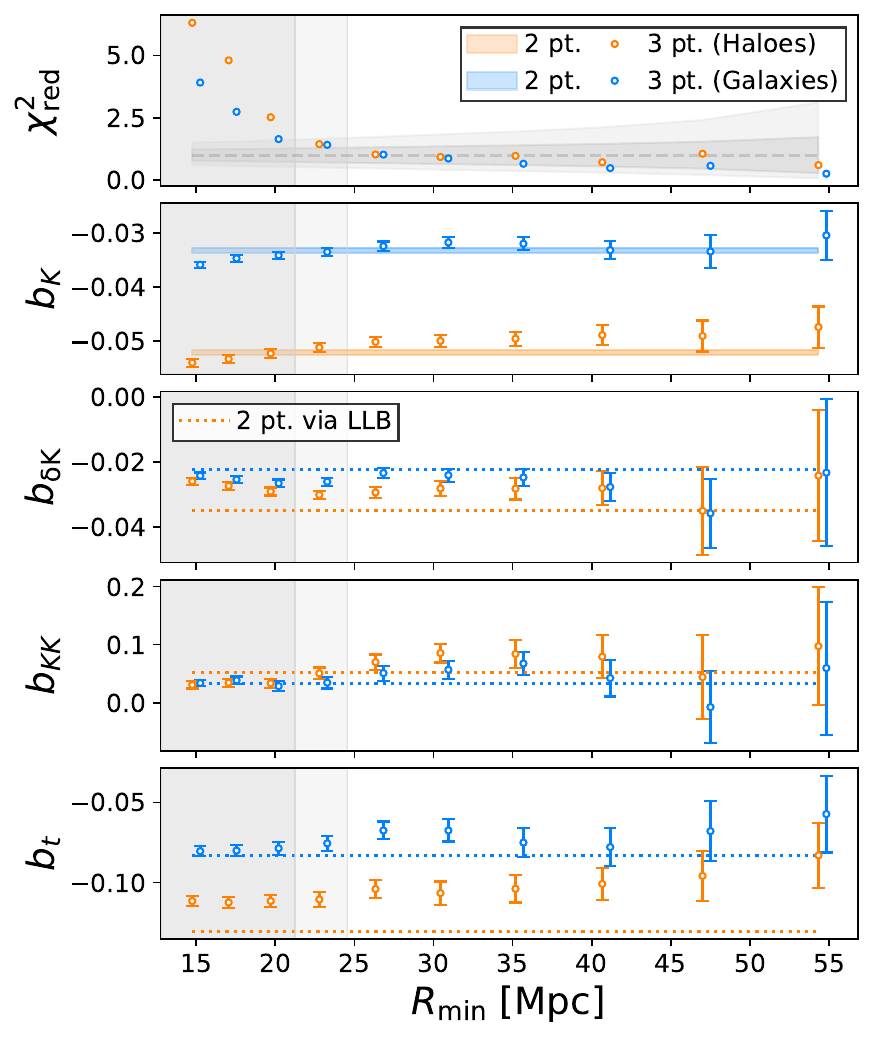}
    \caption{Dependence of the four EFT parameters fitted to $\langle N N M_{\rm ap} \rangle$ on the scale cut $R_{\rm min}$, where all scales with $R_i < R_{\rm min}$ are removed. Here we show our main sample of $M_{\rm h} > 10^{13}$ M$_\odot$ at $z=0$. For the linear bias parameter, the orange and blue shaded regions denote the measurements from two-point statistics, while the shaded region in the $\chi^2_{\rm red}$ panel represents the 1$\sigma$ (2$\sigma$) confidence interval around $\chi^2_{\rm red} = 1$, calculated using the galaxy data vector after the SVD. For the bias parameters, the error bars denote the 68 percent confidence interval. The leftmost grey-shaded region denotes scales whose inclusion would cause $\chi^2_{\rm red}$ to deviate from unity by more than $2\sigma$, while the second grey-shaded region marks the more conservative scale cut used for parameter constraints. The dotted lines show the predictions from the linear Lagrangian ansatz, using the linear bias parameters inferred from two-point statistics.}
    \label{fig:runningparameters}
\end{figure}

We present the measurements for this sample across several aperture radius configurations in Figure \ref{fig:data_fits}, together with the corresponding best-fit EFT predictions obtained using the parameter values given in Eq. \ref{eq:bias_params}. Only a subset of the full data vector is shown to avoid overcrowding. The dependence of the fitted parameters on the minimum scale used in the fit is shown in Figure \ref{fig:runningparameters}.
Overall, we find excellent agreement between the model and the measurements on scales larger than approximately 22 Mpc: the reduced chi-squared is close to unity and the fitted parameters show no significant running with the scale cut. At smaller scales, the fit does not fail abruptly; however, the reduced chi-squared increases noticeably above unity. This effect is most pronounced for equal-scale aperture configurations, where all three $R$'s fall below the scale cut, leading to large deviations from the theoretical predictions.
In this regime, the inferred linear bias parameter, $b_K$, exhibits a slight offset relative to the value obtained from the two-point function, whereas on larger scales the two- and three-point fits are in very good agreement. In an observational setting, the signal-to-noise ratio is expected to be significantly lower, likely rendering this bias statistically insignificant at these scales. At even smaller scales, however, it may become non-negligible. In Figure \ref{fig:runningparameters}, the dotted lines indicate the expectations from the co-evolution relations. While some deviations are present, these relations capture the overall trends reasonably well; we return to this point later.

We note that fitting this model requires aperture mass statistics evaluated at different radii (i.e. different from $R=R_1=R_2=R_3$) in order to break parameter degeneracies. When the analysis is restricted to the equal-scale configuration, the four contributing terms become nearly degenerate, leading to divergent parameter contours. However, each parameter depends differently on the triangle configuration, and incorporating a broader range of configurations can therefore break these degeneracies. Moreover, the equal-scale configuration does not yield the highest SNR; several unequal-scale configurations provide comparable or even higher SNR. This behaviour contrasts with the lensing case for the KiDS-1000 sample, where using only equal-scale apertures does not substantially degrade the signal \citep{Burger2024}. However, as in this work, using different aperture radii can help constrain baryonic effects \citep{burger2025}.

\begin{figure}
    \centering
    \includegraphics[width=\linewidth]{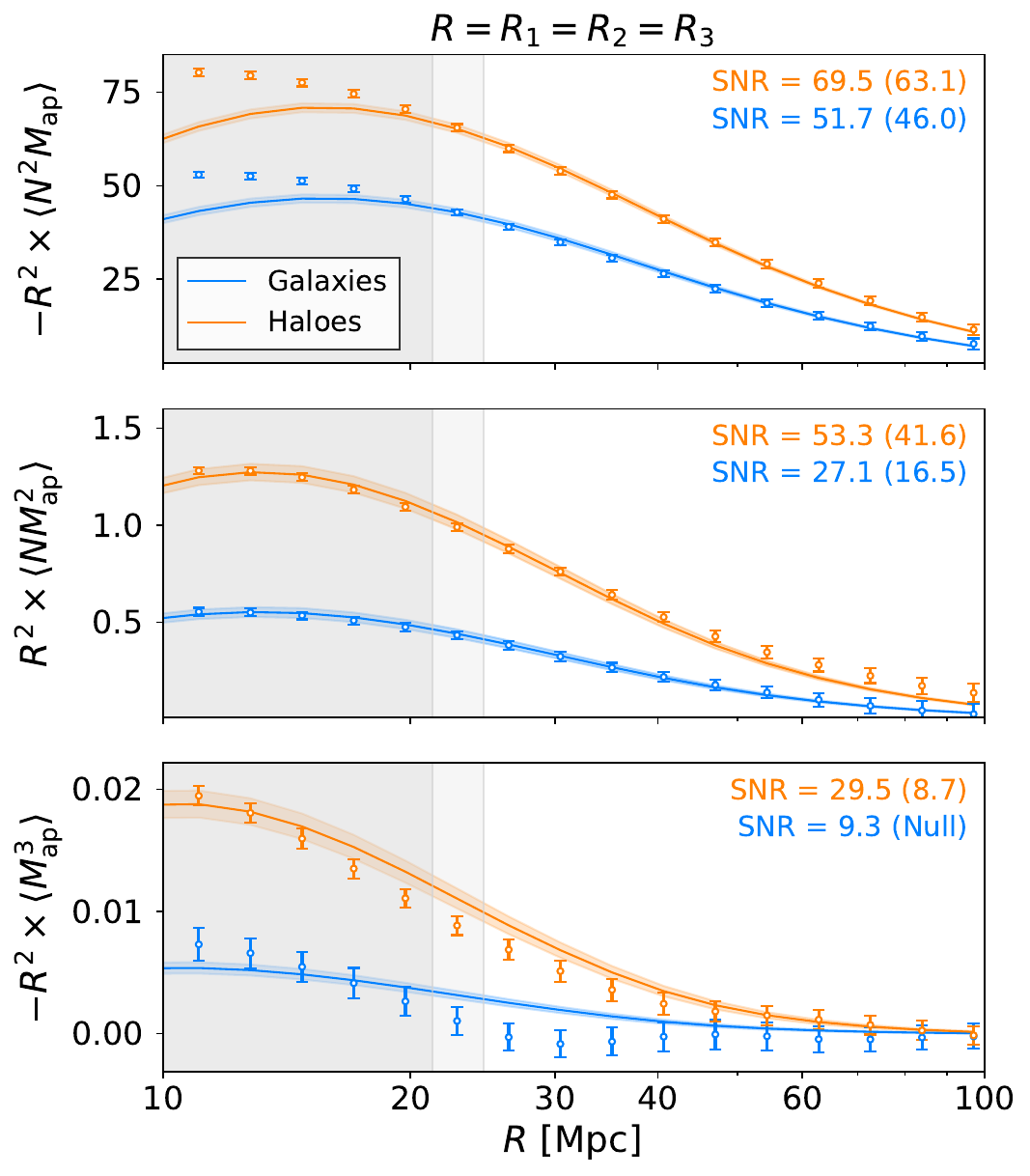}
    \caption{The measured aperture mass signals of our main samples for all three combinations of matter and shapes. The results are presented for equal-scale apertures with $R_1 = R_2 = R_3 = R$.  The corresponding theoretical predictions, using the best fit values from Eq. \ref{eq:bias_params}, are overplotted. The leftmost grey-shaded region marks scales whose inclusion would cause the reduced chi-square, $\chi^2_{\rm red}$, to deviate from unity by more than $2\sigma$, while the second grey-shaded region indicates the more conservative scale cut used for parameter constraints. We emphasize that these predictions are fitted to $\langle N N M_{\rm ap} \rangle$, and therefore \emph{not} fitted to the bottom two panels; instead, we use the fitted parameters and assume zero stochastic noise. For all predictions, the shaded band around the line denotes the 68\% confidence interval. Furthermore, we show the SNR using all aperture radius combinations that satisfy $R_i\geq 26$ Mpc, with in brackets the SNR when using only the equal-scale configurations with the same scale cut.}
    \label{fig:allapertures}
\end{figure}

In Figure \ref{fig:allapertures}, we present the three aperture mass statistics measured for the equal-scale aperture configuration, together with their SNRs. We also show the corresponding SNRs obtained when using all radius configurations with $R_i \geq 26$. The impact of shape noise and misalignment is quite evident. For haloes, the SNR of $\langle N M_{\rm ap}^2 \rangle$ is about 23 percent lower than that of $\langle N N M_{\rm ap} \rangle$, whereas for galaxies the difference is substantially larger. This contrast is even more pronounced for $\langle M_{\rm ap}^3 \rangle$, which yields a much clearer detection for haloes. When restricting the analysis to the equal-scale configuration, this statistic is consistent with a null detection for galaxies on large scales; however, a significant detection is recovered when all configurations are included. This result again highlights the added value of aperture mass statistics measured with different radii. The figure also shows predictions for $\langle N M_{\rm ap}^2 \rangle$ and $\langle M_{\rm ap}^3 \rangle$ derived from parameters fitted solely to $\langle N N M_{\rm ap} \rangle$. We find that these predictions are consistent with the corresponding measurements,\footnote{Although some discrepancies are visible, they remain within the two-sigma level. We caution against relying on visual inspection alone due to strong correlations between data points.} indicating that the inclusion of these additional statistics could further tighten the constraints on the model parameters.

\begin{figure}
    \centering
    \includegraphics[width=\linewidth]{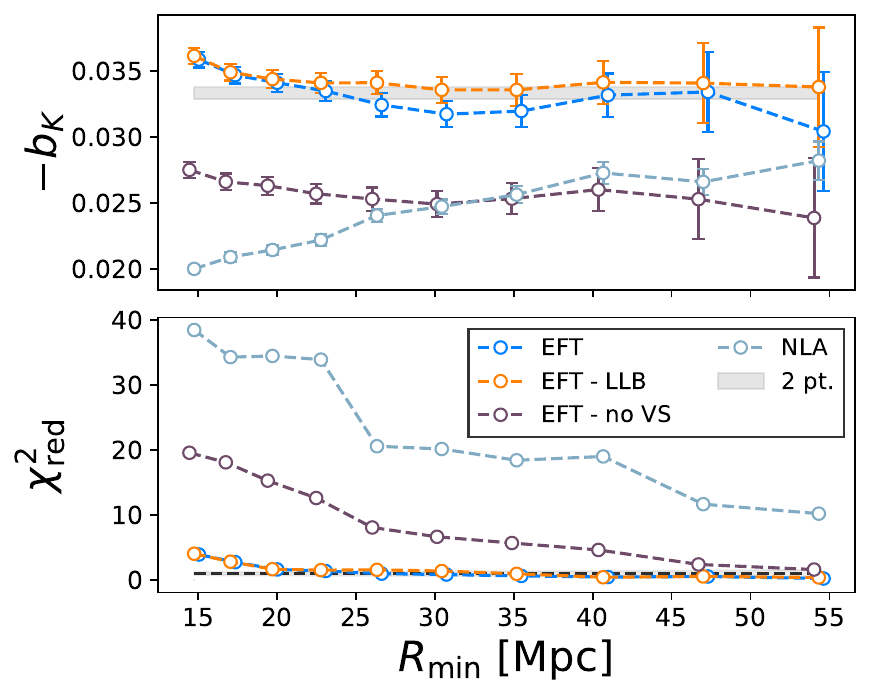}
    \caption{Comparison of several models: EFT, EFT - LLB, EFT - no VS and NLA, all fitted to the $\langle N N M_{\rm ap} \rangle$ data vector. The upper panel shows the measurement of the linear alignment bias coefficient $b_K$ for these models from $\langle N N M_{\rm ap} \rangle$, alongside the measurement from two-point statistics at large scales. The error bars denote the 68 percent confidence interval. The bottom panel shows the reduced chi squared of these models. We show these values as a function of scale cut, where the scale cut is defined as the removal of all scales with $R_i < R_{\rm min}$. Here we show our main galaxy sample of $M_{\rm h} > 10^{13}$ M$_\odot$ at $z=0$. }
    \label{fig:modelcomparison}
\end{figure}

To facilitate a comparison between the EFT framework with other models in the literature, which neglect one or more higher order bias parameters, and to assess the validity of the Lagrangian bias ansatz as a physically motivated approximation for those bias parameters, we contrast the EFT with three alternative models described in Section \ref{sec:modelling}. The results for our main galaxy sample at redshift zero are shown in Figure \ref{fig:modelcomparison}. We consider both the reduced chi-squared, $\chi_{\rm red}^2$, and the bias with respect to the linear amplitude inferred from two-point statistics. Both the full EFT model and the EFT - LLB model, based on the linear Lagrangian bias ansatz, provide accurate descriptions on perturbative scales. In contrast, the phenomenologically motivated EFT - no VS and NLA models fail to capture the behaviour of this sample, even on the largest scales considered. Notably, the EFT - LLB model outperforms the EFT - no VS model despite employing one fewer free parameter. This improvement arises because the co-evolution relations are approximately satisfied and because the velocity–shear operator, \(b_t\), plays an important role, as demonstrated in Figure \ref{fig:tatt_lp_eft_contours}. There we see that the parameters \(b_{\delta K}\) and \(b_{KK}\) take similar values across all three models, while among the higher-order operators it is the velocity–shear term that is detected with the highest significance. This indicates that the poorer performance of the EFT - no VS model stems from the omission of this operator, and that its fit can already be improved by including the co-evolution prediction for this term.

\begin{figure}
    \centering
    \includegraphics[width=\linewidth]{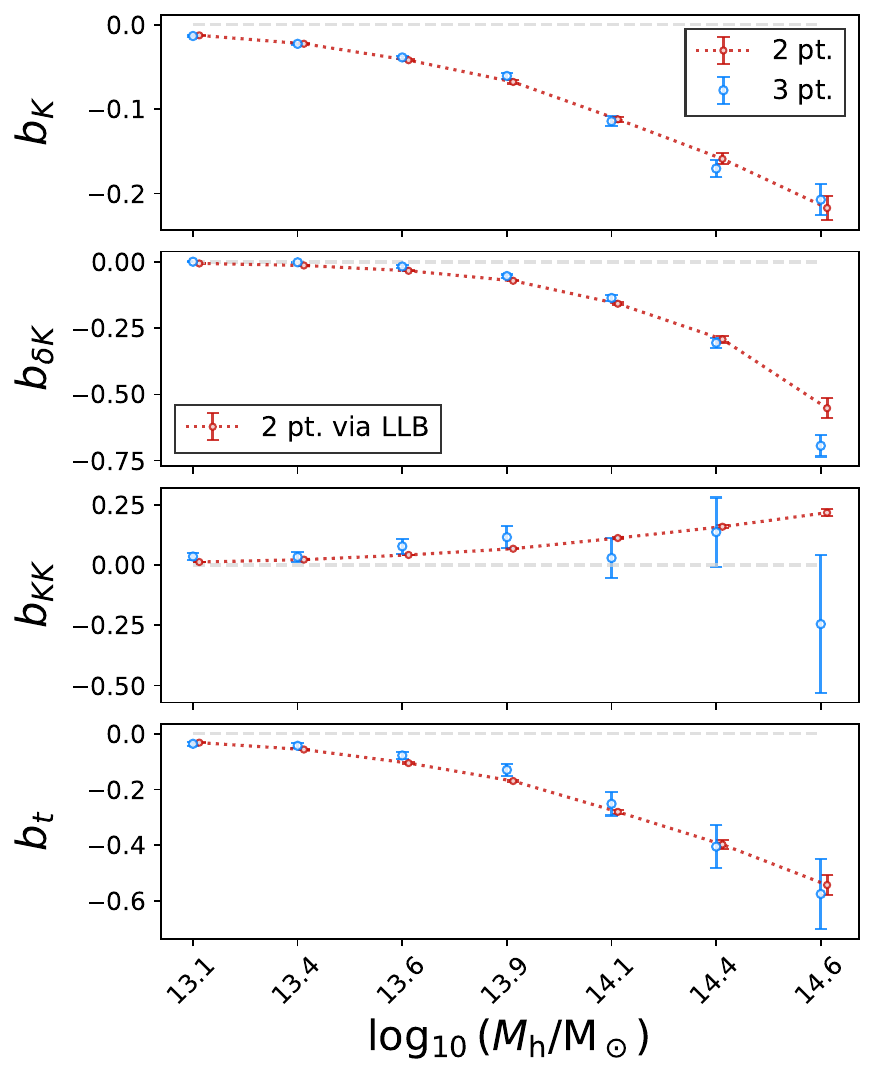}
    \caption{Mass dependence of the EFT parameters for our galaxy sample ($M_{\rm h}>10^{13} \ {\rm M}_\odot$) at $z=0$, where we use a scale cut and omit scales smaller than 26 Mpc. The blue points indicate the three-point measurement using the full EFT. The red points are obtained solely from large scale two-point information: using the linear Lagrangian bias ansatz with the linear alignment bias $b_K$ and the linear galaxy bias $b_1$ measured in the same mass bins. The error bars denote the 68 percent confidence interval.}
    \label{fig:massdependence}
\end{figure}

To study the mass dependence of the bias parameters, we divide our sample into eight halo mass bins spanning the range $10^{13}\,{\rm M}_\odot$ to $10^{15}\,{\rm M}_\odot$. We omit the highest-mass bin, as it contains too little signal to yield meaningful constraints. The corresponding parameter measurements as a function of halo mass are shown in Figure \ref{fig:massdependence}. In each panel, we compare the EFT measurement in that bin with the prediction from the co-evolution relation inferred from $b_K$ and $b_1$ measured using two-point statistics. We emphasize that perfect agreement is not expected; any deviation simply reflects non-linear effects in Lagrangian space. The exception here is the linear parameter, which needs to be in agreement with two-point statistics. We see that this is indeed the case for the complete mass range. For the co-evolution relations the agreement is quite good, similar to the results of \citet{akitsu2023}, which presented these relations for haloes. For all parameters, the co-evolution prediction typically lies within 2$\sigma$ of the three-point measurement. This  explains the good performance of the EFT - LLB model shown in Figure \ref{fig:modelcomparison}. The relative importance of density weighting increases toward higher masses, consistent with the fact that its co-evolution relation depends on both the linear galaxy bias and the alignment bias, each of which grows with halo mass.

\begin{figure*}
    \centering
    \includegraphics[width=1\linewidth]{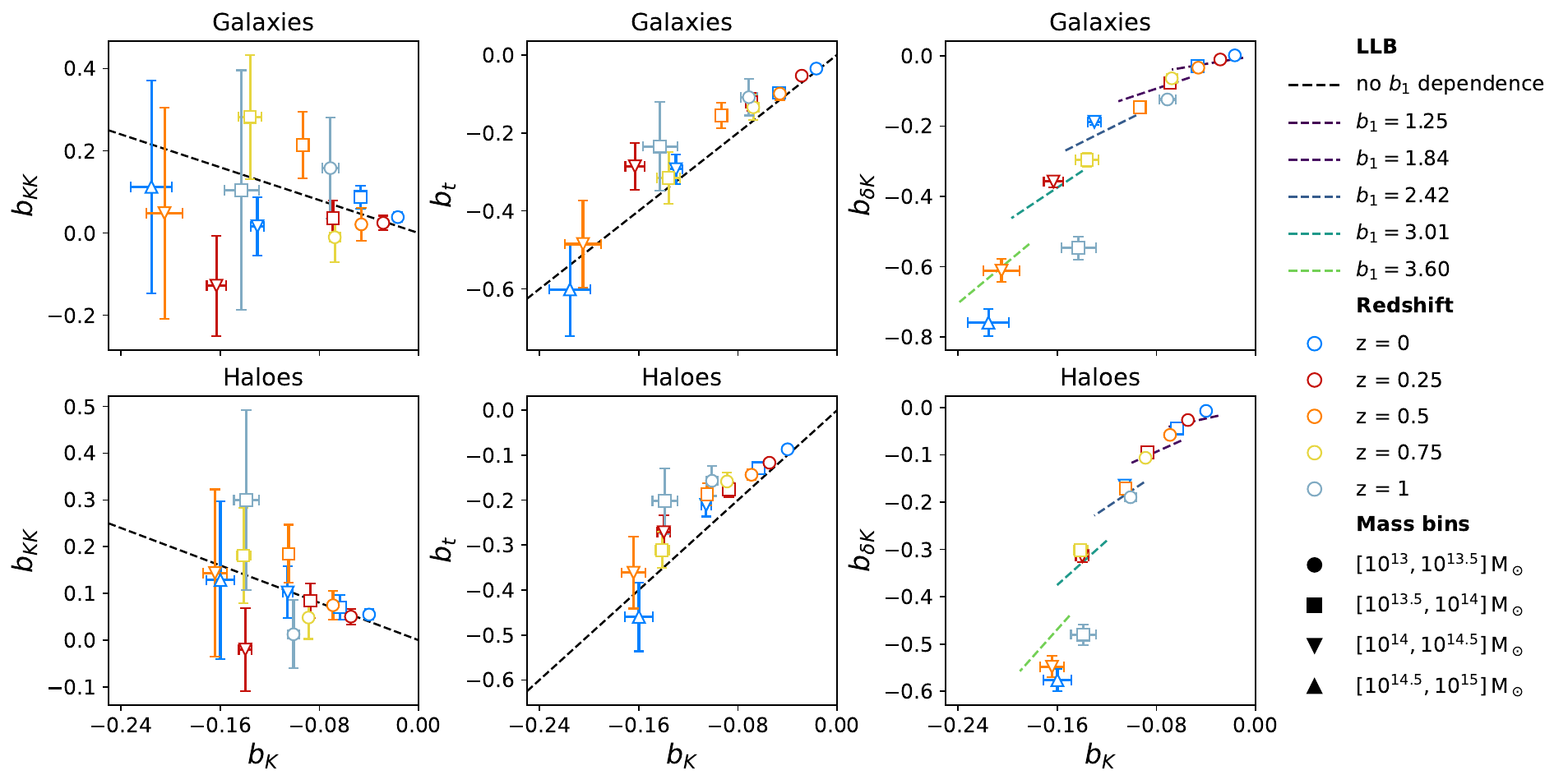}
    \caption{The tidal torquing parameter $b_{KK}$, velocity shear parameter $b_t$ and density weighting parameter $b_{\delta K}$ shown here as functions of the linear bias parameter $b_K$. We show this relation for both galaxies and haloes for 5 redshift bins between 0 and 1, denoted by the different colours and show 4 different halo mass bins between $10^{13}$ and $10^{15} \ {\rm M}_\odot$. All points are obtained with a scale cut removing the points below $26$ Mpc and the error bars denote the 68 percent confidence interval. The dashed lines show the predictions from assuming the linear Lagrangian bias ansatz (co-evolution relations). For the density weighting, this relation depends on both the linear galaxy bias ($b_1$) and the linear alignment bias ($b_K$) of the sample. We illustrate this by plotting several lines that reflect the approximately linear relationship between the two. At higher redshifts, we omit the highest-mass bin(s), as the number of galaxies or haloes satisfying these criteria becomes too small to yield meaningful parameter constraints.  }
    \label{fig:lpfig}
\end{figure*}

This motivates us to examine the generality of these relations across different tracers (galaxies and haloes), redshifts, and masses. We show the relation between all higher order alignment parameters and $b_K$ for 4 halo mass bins between $10^{13}$ and $10^{15} \ {\rm M}_\odot$ and 5 redshift bins between 0 and 1. We present this in Figure \ref{fig:lpfig}, which confirms that the trends identified in Figure \ref{fig:massdependence} are broadly robust. This is again consistent with the findings of \citet{akitsu2023}.
It is worth emphasizing that our analysis differs substantially from previous work: we use full hydrodynamical simulations instead of dark matter only runs, use projected shapes instead of 3D shapes, estimate the alignments in configuration space, and consider both galaxies and haloes. We also remark that we find no significant discrepancy between galaxies and haloes in their adherence to the co-evolution relations, both populations follow the relations with comparable accuracy. Furthermore, we confirmed that these relations are robust to the specific choice of inertia tensor. As discussed in Appendix \ref{app:inertia_tensor}, while the linear alignment amplitude depends on the inertia tensor definition, the higher-order bias parameters follow from a simple rescaling consistent with the co-evolution relations.
Nonetheless, some deviations from the expected relations remain, indicating the presence of non-linear alignment in Lagrangian space. For example, the velocity shear contribution is systematically smaller. This same trend is also apparent in \citet{akitsu2023}. Furthermore, the agreement seems to get worse with both redshift and mass. However, given the sizeable error bars for the higher redshift and mass bins, it is hard to make any definitive statements. Although these relations are not exact, it is important to note that in photometric shear surveys the constraining power on IA parameters is significantly weaker. Under such conditions, the co-evolution relations provide a physically motivated approximation, even if they do not fully capture the statistical precision achieved in the FLAMINGO simulations. We also note that this has implications for the redshift evolution of the higher-order bias parameters, which, to a reasonable approximation, follow a simple evolution tied to the redshift dependence of the linear bias parameters.

Overall, we note that the galaxies follow most of the qualitative features of the haloes, as the higher order bias parameters look similar apart from an overall amplitude rescaling roughly consistent with the co-evolution relations. 

Finally, we find that at the high-mass end the galaxies exhibit a larger alignment amplitude than their host haloes. Although this may seem counter-intuitive given the expected galaxy–halo misalignment, Appendix \ref{app:misalignment} demonstrates that, for these samples, the ellipticity weighting in the estimator dominates over the effect of the misalignment angle. As a result, the galaxies show a higher net alignment amplitude despite being misaligned with their haloes.

\section{Conclusion}\label{sec:concl}
Intrinsic alignments of galaxies constitute a major astrophysical systematic for weak gravitational lensing, and their accurate modelling is essential for the success of forthcoming cosmological surveys. While most existing analyses focus on two-point statistics, higher-order correlations provide access to complementary, non-Gaussian information that is increasingly important in the nonlinear regime. These statistics can significantly improve cosmological parameter constraints by breaking degeneracies between parameters. A consistent theoretical description of these effects is, however, required, as the second order model needs to be compatible with that of third order statistics. Hence, in this work, we focused on the third order alignment statistic. We measured and modelled the third order alignment signal in the largest simulation of the FLAMINGO hydrodynamical simulation suite and compared our results to predictions from several theoretical models.

First, we consider the 3PCF of IA, finding a coherent signal (Figures \ref{fig:stacked_galaxy} and \ref{fig:3pcf}). The aperture mass statistics are then obtained by integrating the 3PCF of matter-shape correlations. We detect the signal for all three third order combinations (Figure \ref{fig:allapertures}). We model $\langle NN M_{\rm ap} \rangle$ using the tree-level EFT of IA and show that this model performs well for both galaxies and haloes for all aperture mass configurations satisfying $R_i \geq 22$ Mpc, giving a reasonable $\chi^2_{\rm red}\approx1$ and consistent results with measurements from two-point statistics (Figures \ref{fig:data_fits} and \ref{fig:runningparameters}). We find that the fitted parameters from $\langle NN M_{\rm ap} \rangle$ are consistent with measurements from $\langle NM_{\rm ap}^2\rangle$ and $\langle M_{\rm ap}^3\rangle$ (Figure \ref{fig:allapertures}). This demonstrates the strength of the EFT in providing a unified and self-consistent description of multiple observables.

We compare the EFT model to several alternatives applied to the same data vector from $\langle NNM_{\rm ap}\rangle$. Specifically, we contrast the full EFT with (i) a version in which the velocity-shear term is set to zero; (ii) a reduced-parameter version in which the velocity-shear and tidal-torquing coefficients are not fitted but instead fixed by the linear Lagrangian bias ansatz (co-evolution relations); and (iii) the NLA model using the \textsc{BiHalofit} bispectrum. The reduced EFT based on the co-evolution relations performs remarkably well. Notably, this reduced EFT model outperforms both the EFT - no VS and NLA models, which yield biased estimates of the linear parameter $b_K$ and give a significantly worse fit, while requiring one fewer parameter than the EFT - no VS model (Figure \ref{fig:modelcomparison}). This highlights the importance of the velocity-shear term and shows that the co-evolution relations capture the higher-order alignment parameters relatively well. This indicates that the `TATT' implementation of \citet{blazek2019} for the power spectrum of IA could potentially be improved by including velocity-shear as well, even if its amplitude is fixed by the co-evolution relations. We further show that the co-evolution relations hold approximately across a wide range of masses and redshifts. However, given our signal-to-noise ratio, we also detect small deviations from these relations, indicating the presence of non-linear alignment already in Lagrangian space (Figure \ref{fig:lpfig}). Finally, we note that galaxies and haloes exhibit no significant qualitative differences in their adherence to these relations. Hence, we conclude that the galaxies broadly follow the alignment properties of their host haloes.

This can be relevant for weak-lensing surveys, where fitting the full EFT parameter set may not be feasible as the constraining power on the alignment parameters will be lower. Our analysis demonstrates that reducing the parameter space by \emph{neglecting} velocity shear, or, in the case of the NLA model, density weighting and tidal torquing as well, can lead to biased constraints on the alignment parameters. This can potentially be problematic for a joint fit. In contrast, adopting the co-evolution relations yields consistent results for the linear parameter and fits the data well, while introducing only a single additional parameter relative to a linear model; and if the galaxy bias of the sample is known, no extra parameters are required at all.

Hence, our results provide a step towards a consistent framework for incorporating higher-order information into two-point analyses. In particular, the EFT and its reduced formulation assuming co-evolution relations provide a self-consistent description of the relevant observables. However, we have also shown that achieving fully consistent results at our signal-to-noise level requires a rather stringent scale cut of $R_i\geq 22$ Mpc. For the purpose of lensing mitigation, this is likely not the optimal strategy. Nevertheless, our findings demonstrate that the EFT framework performs well on large scales. Reaching smaller scales would require extending the model, for example, through a hybrid approach (e.g., \citealt{maion2024}). As stressed before, the utility of third-order statistics lies primarily in their combination with second-order statistics, thus care must be taken to maintain consistency across the various extensions of the model.

In addition, previous studies (e.g., \citealt{sugiyama2024, bakx2025II}) have demonstrated that binning effects and choices of integral cut-offs can significantly influence the results and potentially introduce biases. In our analysis, we find that the minimum integration scale must be chosen to be significantly smaller than the standard cut-off done in the lensing literature, as the higher-order parameters in particular exhibit sensitivity to this choice. In observations, this might not be possible. We therefore emphasize that an explicit and accurate treatment of these effects may be important, and we defer such modelling to future work.

Furthermore, caution is warranted when extrapolating our results to a full cosmological analysis. Due to the resolution limits of the simulations, our study focuses on relatively massive galaxies ($M_{\rm stellar}\gtrsim
 2.2\times10^{11} \ {\rm M}_\odot$), whereas typical weak-lensing samples include many lower-mass systems. Although the EFT formalism is, in principle, agnostic to the details of galaxy formation, as these are absorbed into the bias parameters, it is not obvious a priori that, for example, the Lagrangian ansatz will perform equally well for samples dominated by rotationally supported galaxies. We aim to investigate these questions in future work.

\section*{Acknowledgments}
We thank Mike Jarvis for actively maintaining {\tt TreeCorr}. We thank Joop Schaye for his involvement in the FLAMINGO simulations and for useful comments on this manuscript. CV thanks Lucas Porth, Laila Linke, Rob McGibbon, Aniruddh Herle and Dennis Neumann for several useful discussions. This work has been (partly) funded by the Leiden University Fund. This publication is part of the project ``A rising tide: Galaxy intrinsic alignments as a new probe of cosmology and galaxy evolution'' (with project number VI.Vidi.203.011) of the Talent programme Vidi which is (partly) financed by the Dutch Research Council (NWO). HH acknowledges funding from the European Research Council (ERC) under the European Union's Horizon 2020 research and innovation program (Grant agreement No. 101053992). This work used the DiRAC@Durham facility man-
aged by the Institute for Computational Cosmology on behalf of the STFC DiRAC HPC Facility (www.dirac.ac.uk). The equipment was funded by BEIS capital funding via STFC capital grants
ST/K00042X/1, ST/P002293/1, ST/R002371/1 and ST/S002502/1, Durham University and STFC operations grant ST/R000832/1. DiRAC is part of the National e-Infrastructure.
\section*{Data Availability}
The data supporting the figures in this article are available from the
corresponding author upon reasonable request. The FLAMINGO
simulation data are publicly available \citep{flamingo2026}.\footnote{\hyperref[]{https://flamingo.strw.leidenuniv.nl}}



\bibliographystyle{mnras}
\bibliography{example} 




\appendix

\section{Dependence on inertia tensor}\label{app:inertia_tensor}
\begin{figure*}
    \centering
    \includegraphics[width=1\linewidth]{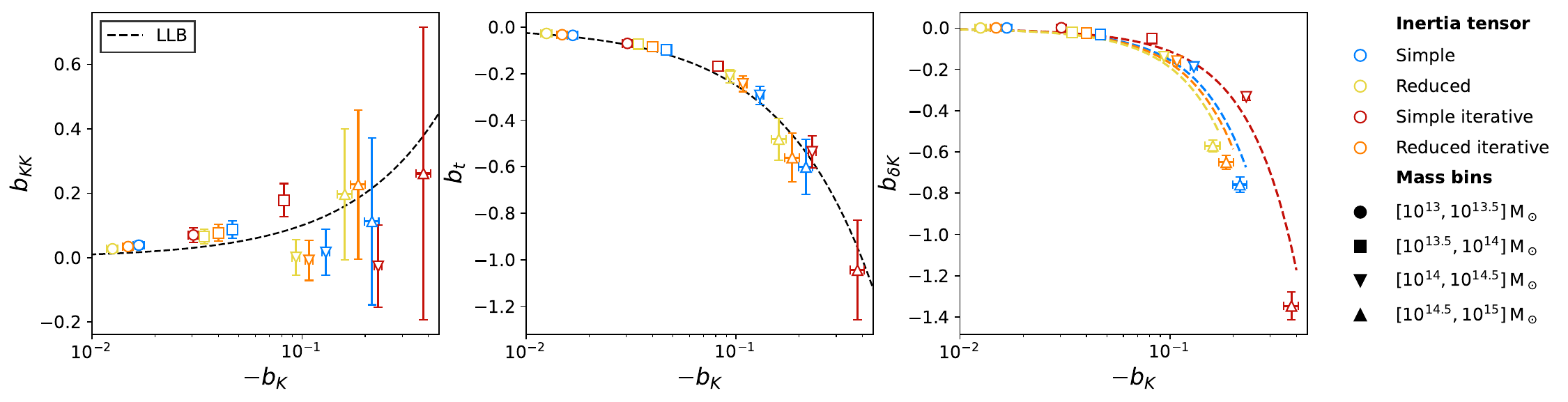}
    \caption{Similar to Figure \ref{fig:lpfig}. The relation between bias parameters for four different inertia tensors at $z=0$: simple, reduced, simple iterative and reduced iterative. Here we show only our galaxy samples. The LLB prediction is over plotted, for the density weighting $b_{\delta K}$, the difference colours show the LLB predictions assuming a linear relation between $b_K$ and $b_1$.}
    \label{fig:lp_fig_shapes}
\end{figure*}
In principle, the bias parameters may depend on the method used to measure galaxy shapes. In this appendix, we examine four different definitions of the inertia tensor. In addition to the simple inertia tensor introduced in Eq. \ref{eq:it}, we consider the reduced inertia tensor, in which particles are downweighted by their 3D distance $r_n$ from the centre,
\begin{equation}
I^{\rm red}_{ij} = \frac{1}{M_{1/2}} \sum_{n=1}^N m^{(n)} \frac{x_i^{(n)} x_j^{(n)}}{r_n^2}.
\end{equation}
Furthermore, we consider an iterative approach. In the iterative approach, the inertia tensor is first computed within a spherical aperture. An equal-volume ellipsoid defined by this tensor is then used to select particles for the next estimate. This is repeated until the axis ratio converges, or 20 iterations are reached, reverting to the initial result if only one particle remains. Combining these choices yields four distinct methods for estimating shapes: simple, simple iterative, reduced, and reduced iterative.
Figure~\ref{fig:lp_fig_shapes} shows the relation between the inferred bias parameters for these different inertia tensor definitions. We display only the galaxy samples, with the LLB prediction over plotted for comparison. 

We note that the linear amplitude varies significantly between different shape measurement methods, with the iterative procedure leading to a larger amplitude and the reduced inertia tensor to a lower one. However, once this amplitude is fixed, the higher-order bias parameters scale in a consistent manner. Consequently, the co-evolution relations are satisfied to a similar degree across all shape measurement methods. Suggesting that these relations are indeed quite general.

\section{Derivation of the line of sight projections for the aperture mass}\label{app:losproj}
We aim to connect the aperture mass statistic to a bispectrum model, e.g. from \citet{bakx2025}. In this work, the bispectrum is expressed in terms of the three momenta $k_1$, $k_2$, $k_3$, and the angles $\mu_1$ and $\xi$ describing the triangles orientation with the line of sight. This decomposition was introduced by \citet{scoccimarro1999}. We briefly summarize this. The quantities $\mu_i$ describe the orientation of each wave vector with respect to the line of sight, while the azimuthal angle $\xi$ specifies a rotation of $\mathbf{k}_2$ around $\mathbf{k}_1$. Without loss of generality, we place $\mathbf{k}_1$ in the $xz$-plane with $k_{1,x} > 0$, which fixes its
direction and allows $\mathbf{k}_2$ to be expressed in terms of $\mu_1$,
$\mu_{12}$, and $\xi$. The remaining vector follows from triangle closure,
$\mathbf{k}_3 = -\mathbf{k}_1 - \mathbf{k}_2$. However, for our purpose it is more convenient to adopt a coordinate system in which each wave vector is expressed as 
$(k_\perp, k_\parallel, \phi)$, where the subscripts $\perp$ and $\parallel$ 
denote components perpendicular and parallel to the line of sight. 
In this basis the three wave vectors are given by
\begin{equation}
\begin{aligned}
\mathbf{k}_1 &= \bigl(k_\perp^{(1)},\, 0,\, k_\parallel^{(1)}\bigr), \\
\mathbf{k}_2 &= \bigl(k_\perp^{(2)} \cos \phi,\, k_\perp^{(2)}\sin\phi,\, k_\parallel^{(2)}\bigr), \\
\mathbf{k}_3 &= -\mathbf{k}_1 - \mathbf{k}_2 .
\end{aligned}
\end{equation}
The $(k_\perp, k_\parallel, \phi)$ basis is related to the $(k,\mu, \xi)$ basis via
\begin{equation}
\begin{aligned}
k_i &= \sqrt{\bigl(k_\perp^{(i)}\bigr)^2 + \bigl(k_\parallel^{(i)}\bigr)^2}, \\[3pt]
\mu_i &= \frac{k_\parallel^{(i)}}{k_i}, \\[3pt]
\cos\xi &= 
\frac{\mu_1 \mu_{12} - \mu_2}
     {\sqrt{(1-\mu_1^{2})(1-\mu_{12}^{2})}},
\end{aligned}
\end{equation}
where $\mu_{12}$ is the cosine of the angle between $\textbf{k}_1$ and $\textbf{k}_2$. The relation above does not uniquely determine $\xi$, since $\sin\xi$ is 
fixed only up to its sign. The correct orientation is obtained by demanding that the sign of $\sin \xi$ is equal to that of  $\hat{\mathbf{n}} \cdot (\mathbf{k}_1 \times \mathbf{k}_2)$, where $\hat{\mathbf{n}}$ is the line-of-sight direction. For completeness, the perpendicular and total magnitudes of $\mathbf{k}_3$ 
follow from triangle closure:
\begin{equation}
\begin{aligned}
k_\perp^{(3)} &= \sqrt{\bigl(k_\perp^{(1)}\bigr)^2 
  + \bigl(k_\perp^{(2)}\bigr)^2
  + 2 k_\perp^{(1)} k_\perp^{(2)} \cos\phi}, \\[4pt]
k_3 &= \sqrt{k_1^2 + k_2^2 
  + 2 k_\parallel^{(1)} k_\parallel^{(2)}
  + 2 k_\perp^{(1)} k_\perp^{(2)} \cos\phi}.
\end{aligned}
\end{equation}

We are now well equipped to calculate $\langle N(R_1)N(R_2)M_{\rm ap}(R_3)\rangle$ in terms of the bispectrum,\footnote{We note that while we focus on position-position-shape, the other correlations can be obtained by interchanging the bispectrum.} starting from Eq. \ref{eq:aperturedef}
\begin{multline}
    \left\langle N(R_1) N(R_2) M_{\rm ap} (R_3)\right\rangle = \\\prod_i^3\int \mathrm{d}^2 \textbf{r}^{(i)}_\perp U_{R_i}\left({r}^{(i)}_\perp\right) \left\langle n\left(\boldsymbol{r}^{(1)}_\perp\right) n\left(\boldsymbol{r}^{(2)}_\perp\right) \epsilon_E\left(\boldsymbol{r}^{(3)}_\perp\right)\right\rangle .
\end{multline}
We consider projected correlations by isolating triplets where the pairs are within a distance $\Pi$ of the shape-galaxy, this can be modelled as follows
\begin{multline}
\left\langle 
n(\boldsymbol{r}^1_\perp)\,
n(\boldsymbol{r}^2_\perp)\,
\epsilon_E(\boldsymbol{r}^3_\perp)
\right\rangle
= 
\int \mathrm{d}\Delta^{(13)}\,\mathrm{d}\Delta^{(23)}
\, W_\Pi(\Delta^{(13)}) W_\Pi(\Delta^{(23)})
\\
\times
\left\langle 
n(\boldsymbol{r}^1)\,
n(\boldsymbol{r}^2)\,
\epsilon_E(\boldsymbol{r}^3)
\right\rangle .
\end{multline}

Here,  we introduced $ r_\parallel^{(2)} = r^{(3)}_\parallel + \Delta^{(23)}$ and $W_\Pi(\Delta)$ is a top hat filter with half-width $\Pi$. Note that here we explicitly used that the shape is located at $\textbf{r}^{(3)}$. Putting all of this together, and inserting in the bispectrum as the inverse Fourier Transform of the term in brackets, we obtain:
\begin{equation}
\begin{aligned}
\langle NN M_{\rm ap}\rangle
&= \prod_{i=1}^3 
   \left[\int \frac{{\rm d}^2 \mathbf{r}^{(i)}_\perp \, {\rm d}^3 \mathbf{k}^{(i)}}{(2\pi)^3}\right]
   \int {\rm d}\Delta^{(13)} {\rm d}\Delta^{(23)} \\[4pt]
&\quad \times 
   U_{R_i}(\mathbf{r}^{(i)}_\perp)\,
   W_\Pi(\Delta^{(13)}) W_\Pi(\Delta^{(23)}) \\[4pt]
&\quad \times (2\pi)^3 
   B_{\delta\delta E}\!\left(
      k_\perp^{(1)}, k_\perp^{(2)}, 
      k_\parallel^{(1)}, k_\parallel^{(2)}, \phi
   \right) \\[4pt]
&\quad \times 
   e^{-i \mathbf{k}_i \cdot \mathbf{r}_i}\,
   \delta\!\left(\mathbf{k}_{1}+\mathbf{k}_{2}+\mathbf{k}_{3}\right),
\end{aligned}
\end{equation}
where we supressed the $R_i$ indices on the left hand side. We now evaluate the Fourier integrals, resulting in 
\begin{equation}
\begin{aligned}
\langle NNM_{\rm ap}\rangle
&= \frac{1}{(2\pi)^6}
   \prod_{i=1}^3 \int {\rm d}^3\mathbf{k}^{(i)}\;
   \tilde{W}_\Pi(k^{(1)}_\parallel)\,
   \tilde{W}_\Pi(k^{(2)}_\parallel) \\[4pt]
&\quad \times 
   \tilde{U}_{R_1}(k_\perp^{(1)})\,
   \tilde{U}_{R_2}(k_\perp^{(2)})\,
   \tilde{U}_{R_3}(k_\perp^{(3)}) \\[4pt]
&\quad \times 
   B_{\delta\delta E}\!\left(
      k_\perp^{(1)}, k_\perp^{(2)},
      k_\parallel^{(1)}, k_\parallel^{(2)}, \phi
   \right) \\[4pt]
&\quad \times 
   \delta\!\left(\mathbf{k}_{1}+\mathbf{k}_{2}+\mathbf{k}_{3}\right).
\end{aligned}
\end{equation}

Evaluating the Dirac delta, enforcing the triangle closure condition $\textbf{k}_{3} = - \textbf{k}_{1} - \textbf{k}_{2}$, then yields
\begin{equation}
\begin{aligned}
\langle N NM_{\rm ap}\rangle
&= \frac{1}{(2\pi)^6}
   \int {\rm d}^3\mathbf{k}^{(1)} {\rm d}^3\mathbf{k}^{(2)}\;
   \tilde{W}_\Pi(k^{(1)}_\parallel)
   \tilde{W}_\Pi(k^{(2)}_\parallel) \\[4pt]
&\quad \times 
   \tilde{U}_{R_1}(k_\perp^{(1)})
   \tilde{U}_{R_2}(k_\perp^{(2)})
   \tilde{U}_{R_3}(||\mathbf{k}^{(1)}_\perp + \mathbf{k}^{(2)}_\perp||) \\[4pt]
&\quad \times 
   B_{\delta\delta E}\left(k_\perp^{(1)}, k_\perp^{(2)}, 
   k_\parallel^{(1)}, k_\parallel^{(2)}, \phi\right).
\end{aligned}
\end{equation}

Finally, we adopt cylindrical coordinates, \ $\mathrm{d}^3\mathbf{k}^{(1)} = k_\perp^{(1)}\mathrm{d}k_\perp^{(1)}\,\mathrm{d}k_\parallel^{(1)}\,\mathrm{d}\phi^{(1)}$. We choose our coordinate system such that $\phi^{(2)} = \phi$, which removes any dependence on $\phi^{(1)}$ in the integrand. We can therefore integrate over $\phi^{(1)}$, yielding
\begin{equation}
\begin{aligned}
\langle N N M_{\rm ap}\rangle
&= \frac{1}{(2\pi)^5}
   \int_0^\infty k_\perp^{(1)}\,{\rm d}k_\perp^{(1)}
   \int_0^\infty k_\perp^{(2)}\,{\rm d}k_\perp^{(2)}  \\
&\quad \times 
   \int_{-\infty}^{\infty} {\rm d}k_\parallel^{(1)}
   \int_{-\infty}^{\infty} {\rm d}k_\parallel^{(2)}
   \int_0^{2\pi} {\rm d}\phi\;
   \tilde{W}_\Pi(k^{(1)}_\parallel)\,
   \tilde{W}_\Pi(k^{(2)}_\parallel) \\[4pt]
&\quad \times 
   \tilde{U}_{R_1}(k_\perp^{(1)})\,
   \tilde{U}_{R_2}(k_\perp^{(2)})\,
   \tilde{U}_{R_3}(||\textbf{k}_\perp^{(1)} + \textbf{k}_\perp^{(2)}||) \\[4pt]
&\quad \times 
   B_{\delta\delta E}\!\left(
      k_\perp^{(1)}, k_\perp^{(2)},
      k_\parallel^{(1)}, k_\parallel^{(2)}, \phi
   \right).
\end{aligned}
\end{equation}
We evaluate these integrals using the {\tt CCL} wrapper \citep{chisari2019} of the {\tt CAMB} linear power spectrum \citep{lewis2000}. The full 5D integral is then evaluated with the {\tt vegas} library \citep{vegas} . The expression above applies to all other combinations of projected positions and shapes as well; only the bispectrum $B_{XYZ}$ and the ordering of the vertices need to be adjusted so that the appropriate triangle vertex is treated as the central one. For example, in position--shape--shape configurations we use $B_{EE\delta}$ in our convention, where the shape tracers are required to lie within a line-of-sight distance $\Pi$ of a central overdensity.
\section{two-point measurements}
\begin{figure}
    \centering
    \includegraphics[width=\linewidth]{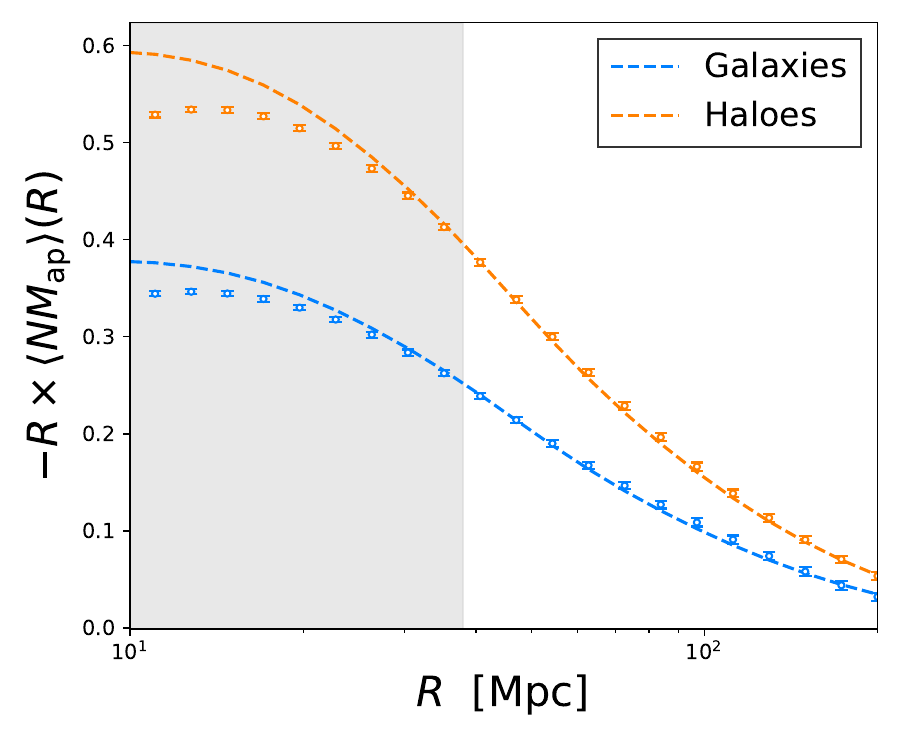}
    \caption{Measurement of $b_K$ from two-point statistics at scales $\geq40$ Mpc, obtained from the two-point aperture mass statistic, for haloes and galaxies with halo mass above $10^{13}$ M$_\odot$ at redshift zero.}
    \label{fig:2pt}
\end{figure}
To obtain an unbiased measurement of the linear galaxy bias $b_1$ and the linear alignment bias $b_K$, we fit simple non linear bias and non linear alignment models to large scales. First, we measure the 2PCF (Eq. \ref{eq:2pcf}) using {\tt TreeCorr}. We then convert these to $\langle N^2\rangle(R)$ and $\langle NM\rangle(R)$ using the standard kernel functions relating the 2PCF to the aperture mass. 

Rather than modelling the aperture mass statistic directly in terms of the power spectrum, which would yield $\langle NM_{\rm ap}\rangle (R)=  \frac{1}{2\pi^2} \int_0^\infty \mathrm{d} k_\parallel \  \mathrm{d} k_\perp \tilde{W}_\Pi(k_\parallel) \tilde{U}^2_R(k_\perp) P_{\delta E}(k)$, as was done for the three-point case, we instead model the second-order aperture mass by modelling the 2PCF and applying the same integrations used for the observational measurements. Although these methods are equivalent on scales well below the upper integration limit, our goal here is to obtain a robust ``ground truth'' for the linear bias parameter. This requires probing the alignment signal on very large scales. Modelling the aperture mass through the 2PCF, rather than directly through the power spectrum, enables us to extend the analysis to these larger scales in a consistent manner.

The 2PCF for a simple non linear alignment model can be related to the non linear power spectrum via \citep{blazek2011} 
\begin{equation}
w_{\rm \delta  \ t}=\frac{b_K}{2\pi^2}  \int_0^\infty \mathrm{d} k_\perp  \mathrm{d} k_\parallel \frac{k_\perp^3}{k^2 k_\parallel} P(k) \sin \left(k_\parallel \Pi\right) J_2\left(k_\perp r_\perp\right),
\end{equation}
and similarly for the clustering 
\begin{equation}
w_{\mathrm{gg}}=\frac{(b_1)^2}{\pi^2} \int_0^{\infty} \mathrm{d} k_\parallel \mathrm{d} k_{\perp} \frac{k_{\perp}}{k_\parallel} P(k) \sin \left(k_\parallel \Pi\right)J_0\left(k_{\perp} r_\perp\right) .
\end{equation}
We convert these to aperture mass statistics using the following integrals,\footnote{These can be derived in a similar fashion to \citet{Schneider2002}.} using the same bins and integration limits as in our measurements
\begin{align}
\langle N M_{\rm ap}\rangle
&= \frac{1}{128}\!\int \!\frac{r_\perp\,{\rm d}r_\perp}{R^2}
   \frac{r_\perp^2}{R^2}\left(12 - \frac{r_\perp^2}{R^2}\right)
   e^{-r_\perp^2/(4R^2)} \\
   &\quad\times\, w_{\rm \delta t}(r_\perp),\nonumber \\
\langle N^2\rangle
&= \frac{1}{128}\!\int \!\frac{r_\perp\,{\rm d}r_\perp}{R^2}
   \left(\frac{r_\perp^4}{R^4} -16\,\frac{r_\perp^2}{R^2} +32\right)
   e^{-r_\perp^2/(4R^2)} \\
&\quad\times\, w_{\rm gg}(r_\perp).\nonumber
\end{align}

Using these, we fit a simple galaxy bias or alignment model on scales above 40 Mpc. Even though it is likely possible to push towards smaller scales, we stress that we are being conservative as our goal is to simply obtain an unbiased measurement of the linear bias parameters. 
For our largest sample, the $b_K$ fits are shown in Figure \ref{fig:2pt}.

\section{B-modes}\label{app:b-mode}

\begin{figure}
    \centering
    \includegraphics[width=\linewidth]{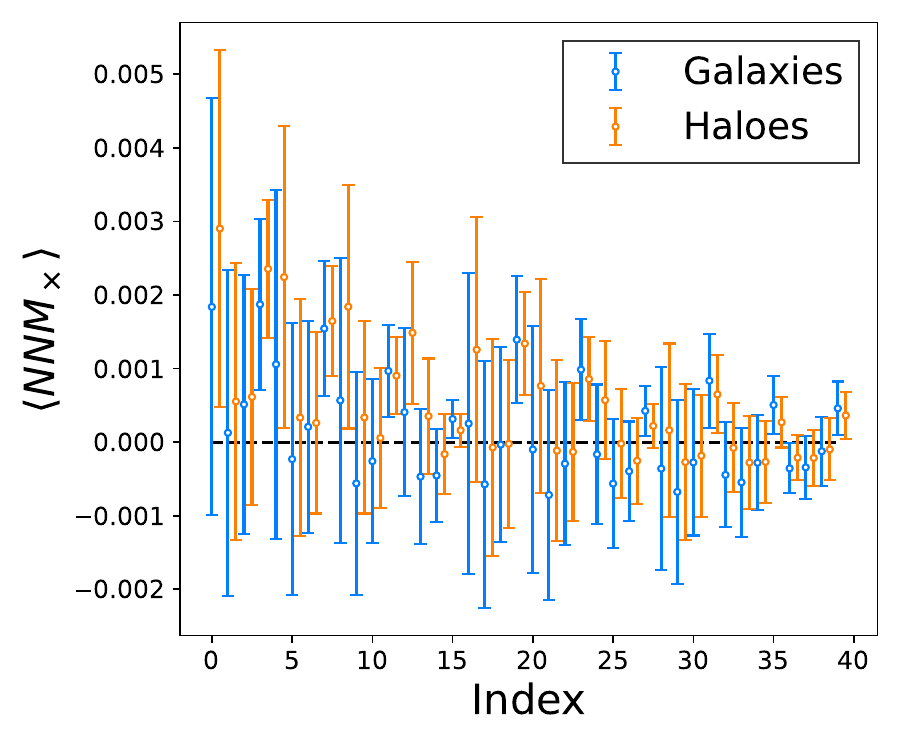}
    \caption{The $\langle NNM_\times \rangle$ statistic for haloes and galaxies with halo mass above $10^{13}$ M$_\odot$ at redshift zero. We consider $R \in [14.76 , 19.71, 26.33, 47.00]$ Mpc. The elements are enumerated in lexicographic order over $(R_1, R_2, R_3)$, with $R_1$ varying slowest and $R_3$ varying fastest. Only combinations satisfying the symmetry condition $R_2 \ge R_1$ are retained, so that redundant (symmetric) points with $R_2 < R_1$ are removed.
Thus the sequence begins $(1,1,1)$, $(1,1,2)$, $(1,1,3)$, $\ldots$, $(1,2,1)$, $(1,2,2)$, $\ldots$.}
    \label{fig:b-mode}
\end{figure}

In \citet{bakx2025II} it was shown that, consistent with EFT predictions, the bispectrum $B_{\delta \delta B}$ can be non-zero. However, we will show in this section that this does not lead to a contribution to $\langle NNM_\times\rangle(R_1, R_2,R_3)$ for any combination of $R$'s. We start with 
\begin{equation}
\begin{aligned}
\langle NN M_{\times}\rangle
&= \frac{1}{(2\pi)^5}
   \int_0^\infty k_\perp^{(1)}\,{\rm d}k_\perp^{(1)}
   \int_0^\infty k_\perp^{(2)}\,{\rm d}k_\perp^{(2)} \\[4pt]
&\quad \times 
   \int_{-\infty}^{\infty} {\rm d}k_\parallel^{(1)}
   \int_{-\infty}^{\infty} {\rm d}k_\parallel^{(2)}
   \int_0^{2\pi} {\rm d}\phi\;
   \tilde{W}_\Pi(k_\parallel^{(1)})\,
   \tilde{W}_\Pi(k_\parallel^{(2)}) \\[4pt]
&\quad \times 
   \tilde{U}_{R_1}(k_\perp^{(1)})\,
   \tilde{U}_{R_2}(k_\perp^{(2)})\,
   \tilde{U}_{R_3}(||\textbf{k}_\perp^{(1)} + \textbf{k}_\perp^{(2)}||) \\[4pt]
&\quad \times 
   B_{\delta\delta B}\!\left(
      k_\perp^{(1)}, k_\perp^{(2)},
      k_\parallel^{(1)}, k_\parallel^{(2)}, \phi
   \right),
\end{aligned}
\end{equation}
where again, we suppressed the arguments on the left hand side. We stress that the bispectrum is non-zero for a range of triplets. However, since the bispectrum of a B-mode is periodic and odd in $\phi$  \citep{bakx2025} its angular average vanishes. We will show this using a short argument, using the short notation $B(\phi) = B_{\delta \delta B}(k^{(1)}_\perp, k^{(2)}_\perp, k^{(1)}_\parallel, k^{(2)}_\parallel, \phi)$ for a fixed set of $k^{(1)}_\perp, k^{(2)}_\perp, k^{(1)}_\parallel, k^{(2)}_\parallel$.
Since $B(\phi)$ is $2\pi$–periodic and odd\footnote{We note that there is also $\phi$ dependence in $\tilde{U}_{R_3}(||\textbf{k}_\perp^{(1)} + \textbf{k}_\perp^{(2)}||)$. However, since this is only via $\cos \phi$ we omit this for brevity.}, we may shift the integration variable by $\pi$ and write
\[
\int_{0}^{2\pi} B(\phi)\,\mathrm{d}\phi
  = \int_{-\pi}^{\pi} B(\phi+\pi)\,\mathrm{d}\phi .
\]
Using oddness and periodicity,
\[
B(-\phi+\pi)
 = -B(\phi-\pi)
 = -B(\phi-\pi+2\pi)
 = -B(\phi+\pi),
\]
so the shifted function is odd on the symmetric
interval $[-\pi,\pi]$.  Hence,
\[
\int_{0}^{2\pi} B(\phi)\,\mathrm{d}\phi
 = \int_{-\pi}^{\pi} B(\phi+\pi)\,\mathrm{d}\phi
 = 0 .
\]
We stress that this is solely due to the angular averaging employed to obtain the aperture mass statistic. Therefore, it might be possible to construct a configuration space estimator that does measure the position-position-shape B-mode. This might be worth pursuing, as there is information in the B-mode that is averaged away for the aperture mass statistic. We test whether this is indeed the case for our fiducial samples in Figure \ref{fig:b-mode} and indeed do not observe a consistent signal. We observe a slight tendency for the signal to be shifted toward positive values; however, this effect is marginal. Moreover, caution is warranted in drawing conclusions from this behaviour, as discrete binning effects can prevent the 3PCF from cancelling exactly, potentially producing such small residuals. This is also consistent with observational results: the KiDS analysis by \citet{Porth2024} likewise finds no evidence for a parity-odd B-mode.

\section{EFT - no VS and EFT - LLB posteriors}
\begin{figure}
    \centering
    \includegraphics[width=\linewidth]{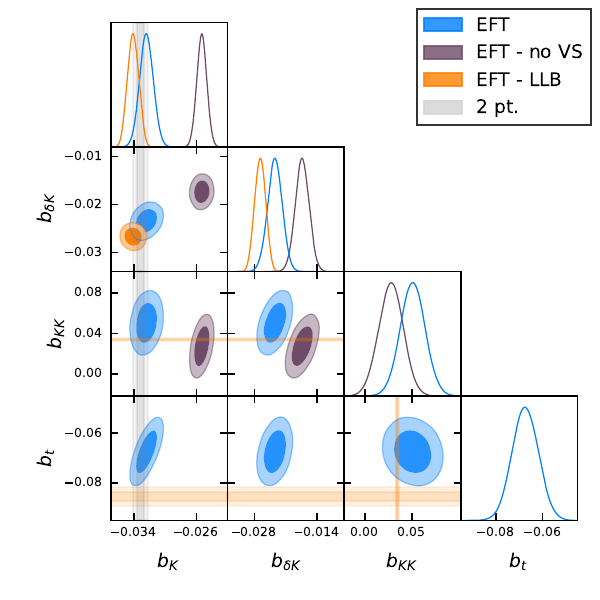}
    \caption{Posteriors of the parameters of the EFT, EFT - no VS and EFT - LLB models fitted to $\langle NNM_{\rm ap} \rangle$ of galaxies at scales $\geq$ 26 Mpc. Here, we show our fiducial sample. The grey lines show the alignment amplitude obtained from two-point statistics at large scales. The one (two) sigma around the two-point estimate is shown as the grey shaded region. Similarly, the orange bars denote the values for $b_{KK}$ and $b_t$ that are used in the EFT - LLB model.}
    \label{fig:tatt_lp_eft_contours}
\end{figure}

This section presents the posterior distributions of the parameters inferred for the EFT, EFT–no VS, and EFT–LLB models fitted to $\langle NNM_{\rm ap} \rangle$.
Figure \ref{fig:tatt_lp_eft_contours} summarizes the constraints for both galaxies and haloes on scales $\geq 26\,\mathrm{Mpc}$. We find that the EFT–no VS model fails to recover the correct linear alignment amplitude, as it neglects the velocity–shear operator $b_t$. In contrast, the EFT–LLB model, which is based on the linear Lagrangian bias ansatz, successfully recovers the correct amplitude with one fewer free parameter by assuming the co-evolution relations for $b_{KK}$ and $b_t$. 

\section{Alignment amplitude: misalignment angle and ellipticity weighting}\label{app:misalignment}
\begin{figure}
    \centering
    \includegraphics[width=\linewidth]{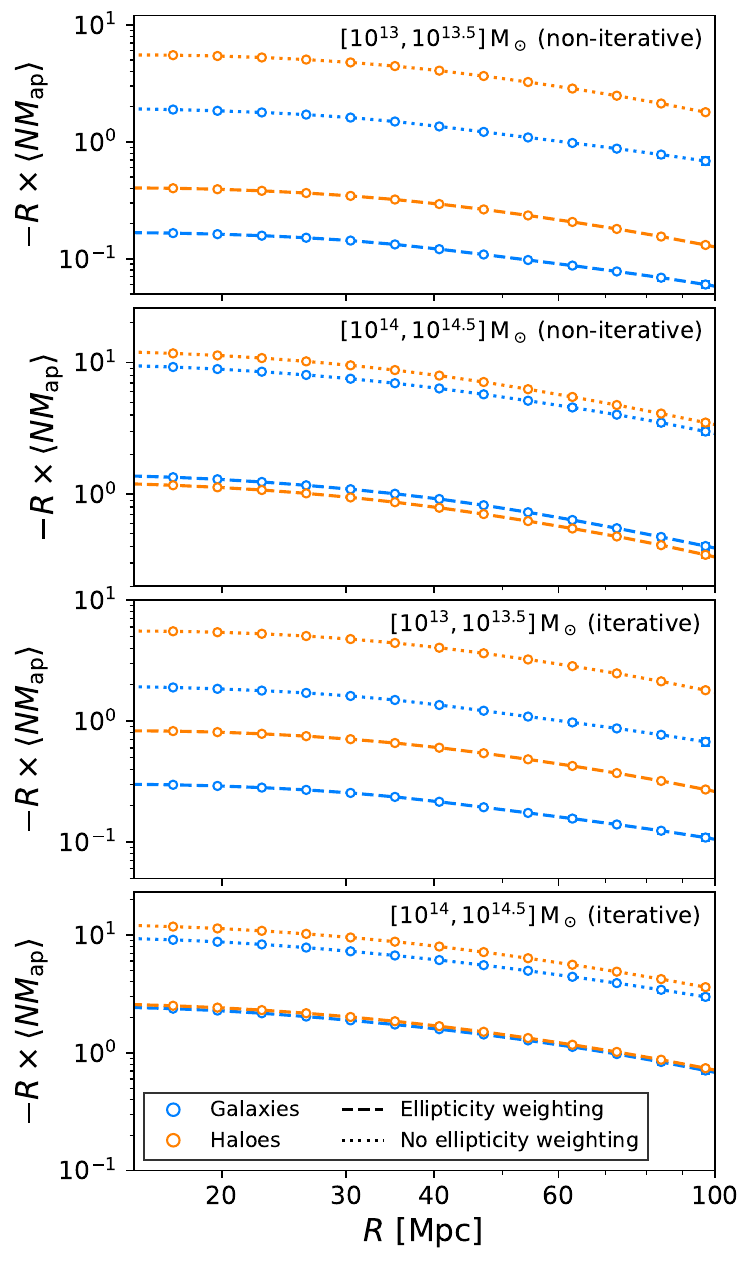}
    \caption{The second-order aperture mass statistics for shape–matter correlations in two different mass bins for two different inertia tensors (simple and iterative). The different line styles denote a difference in ellipticity weighting, the dashed lines indicate our standard estimator, while the dotted lines denote the correlations solely due to the angle. }
    \label{fig:misalignment}
\end{figure}

In Figure \ref{fig:lpfig} we observe that, for more massive systems, the alignment amplitude of the galaxies is higher than that of the haloes they inhabit. At first sight this may seem surprising: the conventional picture is that haloes should exhibit the stronger alignment, because galaxies are typically misaligned with respect to their host halo by some random angle. However, our results do not contradict this expectation. The key point is that the amplitude of the alignment signal depends not only on the orientation angle, but also on the intrinsic ellipticity of the tracer:
\begin{equation}
    \epsilon_1 + i \epsilon_2 = \epsilon_0 \exp( 2i \phi),
\end{equation}
where $\epsilon_0$ is the ellipticity of the galaxy or halo. At high mass, galaxies tend to have smaller misalignment angles relative to their host halo (e.g. \citealt{herle2025}). Hence, the difference in alignment between galaxies and haloes resulting from the misalignment angle will become smaller. Galaxies are also more elliptical than their haloes. Taken together, these effects yield a net increase in the observed alignment amplitude for very massive galaxies relative to their haloes. 

To illustrate this more clearly, Figure \ref{fig:misalignment} presents the second-order aperture mass statistics for shape–matter correlations in two different halo mass bins. We show results for both our fiducial sample and a modified sample in which all ellipticities are normalized to unity. The latter removes ellipticity weighting and isolates the contribution from orientation alone.

We compare measurements based on two definitions of the inertia tensor: the simple non iterative and the simple iterative tensor. For both tensor definitions, the normalized case consistently shows a larger alignment amplitude for haloes than for galaxies, in agreement with the physical expectation that dark matter haloes are more strongly aligned with the surrounding matter distribution.

In the weighted case, however, the larger intrinsic ellipticities of galaxies increase their measured signal. As a result, for the non iterative inertia tensor, galaxies exhibit a larger alignment amplitude in the higher-mass bins compared to the haloes. For the iterative tensor, haloes still have a higher amplitude, but the same ellipticity-weighting effect significantly reduces the difference between haloes and galaxies.

\bsp	
\label{lastpage}
\end{document}